\documentclass[english,twocolumn,a4]{revtex4-1}
\usepackage{amsmath}
\usepackage{graphicx}
\usepackage{amssymb}

\newcommand{\mr}{\mathrm}

\begin{document}

\title{Density response of a trapped Fermi gas: a crossover from the pair vibration mode to the Goldstone mode}

\author{A. Korolyuk$^1$, J. J. Kinnunen$^1$, P. T\"orm\"a$^{1,2}$}
\email{paivi.torma@aalto.fi}
\affiliation{$^1$Department of Applied Physics, School of Science, Aalto University, P.O. Box 15100, 00076 Aalto, Finland\\ $^2$ Kavli Institute for Theoretical Physics, University of California, Santa Barbara, California 93106-4030, USA}

\begin{abstract}
We consider the density response of a trapped two-component Fermi gas. 
Combining the Bogoliubov-deGennes method with the random
phase approximation allows the study of both collective and single particle
excitations. Calculating the density response across a wide range of
interactions, we observe a crossover from a weakly interacting pair 
vibration mode to a strongly interacting Goldstone mode. The crossover
is associated with a depressed collective mode frequency and an increased
damping rate, in agreement with density response experiments performed
in strongly interacting atomic gases.
\end{abstract}

\maketitle

\section{Introduction}

The response of a many-body system to external perturbations can be understood
in terms of collective, i.e. many-body, and single particle excitations.
Collective modes of a quantum fluid are well described by the hydrodynamic
model in the strongly interacting hydrodynamic and the weakly interacting 
collisionless limits~\cite{Griffin1997a,Baranov2000a,Bruun1999a,Stringari2004a,Bulgac2005a}. 
However, experimental studies on collective modes of
trapped Fermi gases~\cite{Kinast2004a,Kinast2005a,Altmeyer2007a,Wright2007a} yielded 
surprises~\cite{Bartenstein2004a,Kinast_hydrodynamics,Altmeyer2007b} that did not fit in the simple picture
obtained from the hydrodynamic theory. The difference was suggested to lie in the 
interplay between the single particle excitations and the collective modes~\cite{Combescot2004a}.

The Bogoliubov-deGennes (BdG) mean-field theory provides a microscopic description of the
trapped Fermi gas. While the single particle excitations are readily accessible from the basic
BdG theory, also the collective modes can be obtained by using the random phase approximation~\cite{Anderson1958} (RPA).
The method allows the study of different modes~\cite{Minguzzi,Bruun_Mottelson,Rodriguez2002a,Bruun_scissors_mode}, but here we will concentrate on the monopole mode in a spherically symmetric trapped Fermi gas.
The method has already been used for studying the monopole mode of a trapped Fermi gas in 
the limits of weak and strong interactions. The purpose of the present work is to study the interesting crossover
region between the weakly interacting and the strongly interacting regimes where the hydrodynamic
theory has failed to properly describe the experiments.

The weakly interacting limit was studied in Ref.~\cite{Bruun2002a},
and in this limit the lowest lying collective mode with the monopole symmetry 
was identified as the pair vibration mode with the frequency $\omega = 2\Delta (0)/\hbar$,
where $\Delta(0)$ is the superfluid excitation gap in the center of the trap. This mode was argued to be the
precursor of the Goldstone mode in the strongly interacting regime with the frequency 
$2\,\omega_{\mr T}$, where $\omega_{\mr T}$ is the harmonic trapping frequency,
in agreement with the strongly interacting hydrodynamic theory. The Goldstone mode in this more strongly 
interacting regime was considered in Ref.~\cite{Grasso2005a}. 

Here we study the actual transition between the two regimes and how the collective 
pair vibration mode transforms into the collective Goldstone mode when the interaction
strength is increased. The crossover region is characterized by a depression of the
collective mode frequencies and increased damping, in qualitative agreement with
experiments done in non-spherically symmetric traps and with various collective 
modes~\cite{Bartenstein2004a,Kinast_hydrodynamics,Altmeyer2007b}.
The full crossover could be studied in future experiments, and the study
of the lately realized  systems with small numbers of atoms~\cite{Serwane2011a} 
might be helpful in finding the required parameters.

This paper is organised as follows. In Section~\ref{sec:linear_response} we review the standard
theory of linear response for a small disturbance to the system and
discuss the connection between the density response and the frequency
of collective excitations. In Section~\ref{sec:method} we discuss the details of the
methods used. In Section~\ref{sec:results} we present and analyse our results
and Section~\ref{sec:discussion} summarises the main findings and conclusions of this work.

\section{Linear response and collective behaviour}
\label{sec:linear_response}

In this section we review the calculation showing the connection between
the density response and the frequency of collective excitations of
a system. This shows how peaks in the imaginary part of the density
response function correspond to the frequencies of collective excitations,
providing the theoretical background for interpreting the results
of the numerically calculated density response function.

\subsection{The density response function \label{sub:Density-response-function}}

Let us consider a system described by a full many-body Hamiltonian
$\hat{H}$. It is assumed that the system is initially in a pure state
$\left|\phi_{0}\right\rangle $ which is an eigenstate of the Hamiltonian
$\hat{H}$, for example the ground state. At the moment $t=0$
a probing field or a perturbation $\hat{V}$ is switched on. From
that moment the system evolves under the Hamiltonian $\hat{H'}=\hat{H}+\hat{V}$.
We denote the state of the system at the time $t$ as $\left|\phi(t)\right\rangle $.
In the absence of any probing field, the state of the system at time
$t$ would remain $\left|\phi_{0}\right\rangle $. The difference
between $\left|\phi(t)\right\rangle $ and $\left|\phi_{0}\right\rangle $
is thus caused by the probing field $\hat{V}$. This difference can
be measured with the help of an observable $\hat{O}(\mathbf{r})$.
Without the probing field, the average of the operator $\hat{O}(\mathbf{r})$
(the statistical average after a series of measurements) would be
\begin{equation}
O_{0}(\mathbf{r})=\left\langle \phi_{0}\right|\hat{O}(\mathbf{r})\left|\phi_{0}\right\rangle .\label{eq:A_0}\end{equation}
 When the field $\hat{V}$ is switched on, the measurements would
yield \begin{equation}
O(\mathbf{r},t)=\left\langle \phi(t)\right|\hat{O}(\mathbf{r})\left|\phi(t)\right\rangle .\label{eq:A}\end{equation}
The difference 
\begin{equation}
  \delta O(\mathbf{r},t)=O(\mathbf{r},t)-O_{0}(\mathbf{r})
  \label{eq:dA}
\end{equation}
indicates how the perturbation $\hat{V}$ has influenced the physical
observable $\hat{O}$.

The difference, or response, $\delta O$ can be calculated in the
interaction picture representation.
In the density response
approach we consider $\hat{V}$ as a weak disturbance to the system.
In this case we can take into account only the first (linear) order
of the perturbation $\hat{V}$. Thus the response is
\begin{equation}
  \delta O(\mathbf{r},t)\approx-i\intop_{-\infty}^{t}dt'\left\langle \phi_{0}\right|\left[\hat{O_{\mr I}}(\mathbf{r},t),\hat{V_{\mr I}}(t')\right]\left|\phi_{0}\right\rangle,
  \label{eq:dA_long}
\end{equation}
where $\hat{O_{\mr I}}$ and $\hat{V_{\mr I}}$ are the operators $\hat{O}$
and $\hat{V}$ in the interaction picture representation.

In this work we are interested in the particular case of an external
potential $\upsilon(\mathbf{r},t)$ that couples to the density $\hat{\rho}(\mathbf{r})$,
corresponding to an operator $\hat{V}=\int d\mathbf{r}\hat{\rho}(\mathbf{r})\upsilon(\mathbf{r},t)$.
Then $\delta O$ becomes
\begin{equation}
  \delta O(\mathbf{r},t)=\intop_{-\infty}^{+\infty}dt'd\mathbf{r}'\mathcal{A}(\mathbf{r},\mathbf{r}',t,t')\upsilon(\mathbf{r}',t'),
  \label{eq:dA_khi}
\end{equation}
where the kernel
\begin{equation}
  \mathcal{A}(\mathbf{r},\mathbf{r}',t,t')=-i\left\langle \phi_{0}\right|\left[\hat{O_{\mr I}}(\mathbf{r},t),\hat{\rho_{\mr I}}(\mathbf{r}',t')\right]\left|\phi_{0}\right\rangle \theta\left(t-t'\right)
  \label{eq:khi_A}
\end{equation}
is called the response function.

The response function $\mathcal{A}(\mathbf{r},\mathbf{r}',t,t')$ in
Eq.~\eqref{eq:khi_A} shows the change in the observable $\hat O$
measured at the point $\left(\mathbf{r},t\right)$ due to the infinitesimally
small perturbation at the point $\left(\mathbf{r}',t'\right)$.
If the Hamiltonian $\hat{H}$ does not depend on time, then 
the time-dependence of $\mathcal{A}$ will be on the 
difference $t-t'$ only.

The derivation above holds for any general observable $\hat{O}$.
In the special case $\hat{O}=\hat{\rho}$, i.e. when the observed
quantity is also the density, the response function is called the
density response function, and the expression for it is
\begin{equation}
  \mathcal{A}(\mathbf{r},\mathbf{r}',t-t')=-i\left\langle \phi_{0}\right|\left[\hat{\rho_{\mr I}}(\mathbf{r},t),\hat{\rho_{\mr I}}(\mathbf{r}',t')\right]\left|\phi_{0}\right\rangle \theta\left(t-t'\right).
  \label{eq:khi_rho}
\end{equation}

\subsection{The frequency of the collective excitations in the density response
\label{sub:Frequency-of-collective}}

After a Fourier transformation of Eq.~\eqref{eq:khi_rho}
one obtains
\begin{equation}
  \mathcal{A}(\mathbf{r},\mathbf{r}',\omega)=-i\intop_{-\infty}^0 dt\,e^{-i\omega t}\left\langle \phi_0\right|\left[\hat{\rho_{\mr I}}(\mathbf{r},0),\hat{\rho_{\mr I}}(\mathbf{r}',t)\right]\left|\phi_0\right\rangle.
  \label{eq:fourier_response}
\end{equation}

We assume, formally, that the Hamiltonian $\hat{H}$ is diagonalized,
and $E_{n}$ and $\left|n\right\rangle $ are its eigenvalues and
eigenvectors: $\hat{H}\left|n\right\rangle =E_{n}\left|n\right\rangle $,
$n=0,1,\ldots$. As eigenstates of a hermitian operator, the eigenvectors
are orthogonal $\left\langle n\right|\left.n'\right\rangle =\delta_{n,n'}$
and form a complete basis $\sum_{n}\left|n\right\rangle \left\langle n\right|=\hat{1}$.
One can assume that the initial state of the system, earlier denoted
as $\left|\phi_{0}\right\rangle $, is the ground state $\left|0\right\rangle $:
$\left|\phi_{0}\right\rangle \equiv\left|0\right\rangle $.

Transforming back to the Schr\"odinger picture representation $\hat{\rho_{\mr I}}(\mathbf{r},t)=e^{i\hat{H}t}\hat{\rho}e^{-i\hat{H}t}$, Eq.~\eqref{eq:fourier_response} becomes 
\begin{equation}
\begin{split}
  \mathcal{A}(\mathbf{r},\mathbf{r}',\omega)&=2  \intop_{-\infty}^{0}dt \, e^{-i\omega t}  \\& \operatorname{Im} \left(e^{-iE_0t} \langle \phi_0\right| \hat{\rho}(\mathbf{r})e^{i\hat{H}t}\hat{\rho}(\mathbf{r}')\left|\phi_0 \rangle  \right) 
  \label{eq:app-1}
\end{split}
\end{equation}
Using the completeness of the eigenbasis and performing the time integral 
yields
\begin{equation}
\begin{split}
  \mathcal{A}(\mathbf{r},\mathbf{r}',\omega)&=\sum_{n}\frac{\left\langle \phi_{0}\right|\hat{\rho}(\mathbf{r})\left|n\right\rangle \left\langle n\right|\hat{\rho}(\mathbf{r}')\left|\phi_{0}\right\rangle }{\omega-\left(E_{n}-E_{0}\right)}\\
&-\frac{\left\langle \phi_{0}\right|\hat{\rho}(\mathbf{r}')\left|n\right\rangle \left\langle n\right|\hat{\rho}(\mathbf{r})\left|\phi_{0}\right\rangle }{\omega+\left(E_{n}-E_{0}\right)}.
  \label{eq:khi_differ}
\end{split}
\end{equation}

Thus, the density response $\mathcal{A}(\mathbf{r},\mathbf{r}',\omega)$
yields information about the excitation spectrum of the system $E_{n}-E_{0}$.
In Section~\ref{sec:results} we will analyse the numerically calculated
density response function and identify the poles of Eq.~\eqref{eq:khi_differ},
i.e. the frequencies $\omega$ in which $\mathcal{A}$ has peaks,
as the collective or single particle excitation frequencies. Note
that here, formally, $E_{n}$ includes all excitations of the system,
both collective and single particle ones.

\section{The Random Phase Approximation for the harmonic trap geometry}
\label{sec:method}

In the previous section we discussed the general definitions of density
response function and its connection to the collective excitation
frequencies. Here we will apply the random phase approximation (RPA)
\cite{Cote_Griffin,Anderson1958} to express the density response
using Green's functions. These in turn will be solved using the Bogoliubov-deGennes
(BdG) mean-field theory, allowing an efficient solution of the density
response for a spherically symmetric trapped atomic gas.

\subsection{Density response expressed using Green's functions \label{sub:Density-response-and}}

A dilute ultracold two-component atomic gas trapped in
a spherically symmetric trap can be described with the Hamiltonian
\begin{equation}
  \hat{H}=\hat{K}+\hat{U},
  \label{eq:Hamiltonian}
\end{equation}
where 
\begin{equation}
  \hat{K}=\sum_{\alpha=\left\{ \uparrow,\downarrow\right\} }\int d\mathbf{r}\psi_{\alpha}^{\dagger}(\mathbf{r})\Bigl[\frac{-\nabla^{2}}{2m}-\mu+\frac{m\omega_{\mr T}r^{2}}{2}\Bigr]\psi_{\alpha}(\mathbf{r})
  \label{eq:K}
\end{equation}
is the single particle Hamiltonian containing the kinetic energy
and the harmonic trapping potential of frequency $\omega_{\mr T}$, $\psi_{\alpha}^{(\dagger)}({\bf r})$
is the annihilation (creation) operator of an atom in a
state $\alpha$ at point ${\bf r}$, and 
\begin{equation}
  \hat{U}=\frac{1}{2}\sum_{\alpha,\beta=\left\{ \uparrow,\downarrow\right\} }\int d\mathbf{r}d\mathbf{r}'\psi_{\alpha}^{\dagger}(\mathbf{r'})\psi_{\beta}^{\dagger}(\mathbf{r}')\psi_{\beta}(\mathbf{r})_{\alpha}\psi(\mathbf{r})g\left(\mathbf{r}'-\mathbf{r}\right)
  \label{eq:U-prev}
\end{equation}
 describes two-body interactions. The short-range interactions can
be described by the two-body scattering T-matrix in the contact potential
approximation $g\left(\mathbf{r}'-\mathbf{r}\right)=g_{0}\delta\left(\mathbf{r}'-\mathbf{r}\right)$,
where the coupling constant $g_{0}$ is related to the physical s-wave
scattering length through the relation $1/g_{0}=\frac{m}{4\pi a}-\sum_{k}\frac{1}{2\epsilon_{k}}$~\cite{Giorgini2008a}.
Eq.~\eqref{eq:U-prev} becomes 
\begin{equation}
  \hat{U}=\frac{g_{0}}{2}\sum_{\alpha,\beta=\left\{ \uparrow,\downarrow\right\} }\int d\mathbf{r}\psi_{\alpha}^{\dagger}(\mathbf{r})\psi_{\beta}^{\dagger}(\mathbf{r})\psi_{\beta}(\mathbf{r})\psi_{\alpha}(\mathbf{r}).
  \label{eq:U}
\end{equation}
For the calculation of the density response we need to add a weak
probing field ${\hat V}$ (for more details see~\ref{sub:Density-response-function}).
We consider 
\begin{equation}
\begin{split}
  {\hat V}= \int d\mathbf{r} &\left[ \phi_{\uparrow}(\mathbf{r},t) n_{\uparrow}(\mathbf{r})+ \phi_{\downarrow}(\mathbf{r},t) n_{\downarrow}(\mathbf{r}) \right.\\
&\left.+\eta (\mathbf{r},t) \psi_{\downarrow}(\mathbf{r})\psi_{\uparrow}(\mathbf{r})+H.c.\right].
  \label{eq:Perturbation}
\end{split}
\end{equation}
The three terms in ${\hat V}$ allow the calculation of the density
responses to the three external fields $\phi_{\uparrow}(\mathbf{r},t),\phi_{\downarrow}(\mathbf{r},t),\eta(\mathbf{r},t)$,
which couple to the system in different ways. To simplify the notation,
we consider a response to a generic field $h(\mathbf{r},t)$, where
$h(\mathbf{r},t)$ can be any of these three external fields.

As physical observables for which we will analyse the response, we
consider the densities $\rho_{\uparrow}$ and $\rho_{\downarrow}$
of up and down components, respectively. Additionally, we consider
the response of the order parameter $\Delta (\mathbf{r})$ to the probing field.
We expect that the same frequencies of collective excitations will
appear in all three responses; naturally, this can be confirmed numerically.

A useful way of formulating the density response is by using the two-point
Green's functions $G_{ij}(\mathbf{1},\mathbf{2})=-\Bigl\langle T\Psi_{i}(\mathbf{1})\Psi_{j}^{\dagger}(\mathbf{2})\Bigr\rangle$
in the Nambu formalism, where ${\bf 1}$ denotes space-time point
$\mathbf{x}_{1},t_{1}$ and 
$\Psi(\mathbf{1})=\left[\begin{array}{c}
\psi_{\uparrow}(\mathbf{1})\\
\psi_{\downarrow}^{\dagger}(\mathbf{1})\end{array}\right]$.
The Green's function for $\mathbf{1}=\mathbf{2}$ has a simple physical
meaning: 
\begin{equation}
  \widehat{G}(\mathbf{1},\mathbf{1})=\left(\begin{array}{cc}
    \rho_{\uparrow}(\mathbf{1}) & \Delta(\mathbf{1})\\
    \Delta^{*}(\mathbf{1}) & -\rho_{\downarrow}(\mathbf{1})\end{array}\right),
  \label{eq:a-1}
\end{equation}
 where $\rho_{\uparrow}(\mathbf{1}),\rho_{\downarrow}(\mathbf{1})$
are the densities and $\Delta(\mathbf{1})$ is the order parameter,
i.e. the gap at the point $\mathbf{1}=\left(\mathbf{x}_{1},t_{1}\right)$.

We consider the response of the Green's function:
\begin{equation}
  \delta\hat{G}_{ij}(\mathbf{1},\mathbf{2})=\intop_{-\infty}^{+\infty}dt_{5}\int_{V}d\mathbf{x}_{5}A_{ij}(\mathbf{1},\mathbf{2},\mathbf{5})h(\mathbf{5}).
  \label{eq:dA_khi-1-1}
\end{equation}
This can be interpreted as the change of the Green's function, or
how the transition between the points $\left(\mathbf{x}_{1},t_{1}\right)$
and $\left(\mathbf{x}_{2},t_{2}\right)$ alters due to a probing field
$h$ at the point $(\mathbf{x}_{5},t_{5})$. The kernel can be written
as
\begin{equation}
  A{}_{ij}(\mathbf{1},\mathbf{2},\mathbf{5})=\left.\frac{\delta G_{ij}(\mathbf{1},\mathbf{2})}{\delta h(\mathbf{5})}\right|_{h\rightarrow0}.
  \label{eq:density_response}
\end{equation}

For the one-point Green's functions the response is
\begin{equation}
  \delta\hat{G}_{ij}(\mathbf{1},\mathbf{1})=\intop_{-\infty}^{+\infty}dt_{5}\int_{V}d\mathbf{x}_{5}\widetilde{A}_{ij}(\mathbf{1},\mathbf{5})h(\mathbf{5}),
  \label{eq:dA_khi-1-1-1}
\end{equation}
where
\begin{equation}
  \widetilde{A}_{ij}(\mathbf{1},\mathbf{3})\equiv A_{ij}(\mathbf{1},\mathbf{1},\mathbf{3})
  \label{eq:A_tilda}
\end{equation}
and
\begin{equation}
  \widetilde{A}{}_{ij}(\mathbf{1},\mathbf{5})=\left.\frac{\delta G_{ij}(\mathbf{1},\mathbf{1})}{\delta h(\mathbf{5})}\right|_{h\rightarrow0}.
  \label{eq:density_response-1}
\end{equation}

As in Eq.~\eqref{eq:a-1}, this is a 2x2 matrix with the different
elements describing the response to different fields. For example,
the response of the density of up particles $\widetilde{A}_{\uparrow\uparrow}(\mathbf{1},\mathbf{5})$
is
\begin{equation}
  \delta\rho_{\uparrow}(\mathbf{1})=\intop_{-\infty}^{+\infty}dt_{5}\int_{V}d\mathbf{x}_{5}\widetilde{A}_{\uparrow\uparrow}(\mathbf{1},\mathbf{5})h(\mathbf{5}).
  \label{eq:dA_khi-1-1-1-1}
\end{equation}

Starting from the Hamiltonian \eqref{eq:Hamiltonian} and using the
random-phase approximation (RPA) \cite{Cote_Griffin}, we write (for
more details see Appendix~\ref{sec:Density-response-within}) 
\begin{widetext}
\begin{equation}
  A_{ij}(\mathbf{1},\mathbf{2},\mathbf{5})= A_{0ij}(\mathbf{1},\mathbf{2},\mathbf{5})+g_{0}\sum_{k,l}\int d\mathbf{3}L_{ikkj}(\mathbf{1},\mathbf{2},\mathbf{3})A_{ll}(\mathbf{3},\mathbf{3},\mathbf{5})-g_{0}\sum_{k,l}\int d\mathbf{3}L_{iklj}(\mathbf{1},\mathbf{2},\mathbf{3})A_{kl}(\mathbf{3},\mathbf{3},\mathbf{5}),
  \label{eq:Main}
\end{equation}
 where $L_{iklj}(\mathbf{1},\mathbf{2},\mathbf{3})=G_{ik}(\mathbf{1},\mathbf{3})G_{lj}(\mathbf{3},\mathbf{2})$,
and $A_{0}(\mathbf{1},\mathbf{2},\mathbf{5})=L_{i\uparrow\uparrow j}(\mathbf{1},\mathbf{2},\mathbf{5})$,
$L_{i\downarrow\downarrow j}(\mathbf{1},\mathbf{2},\mathbf{5})$,
$L_{i\uparrow\downarrow j}(\mathbf{1},\mathbf{2},\mathbf{5})+L_{i\downarrow\uparrow j}(\mathbf{1},\mathbf{2},\mathbf{5})$
for $h=\phi_{\uparrow}$, $\phi_{\downarrow}$, $\eta$ respectively. 
For the density response $\widetilde{A}_{ij}(\mathbf{1},\mathbf{5})$
the equation in Eq.~\eqref{eq:Main} becomes
\begin{equation}
  \widetilde{A}_{ij}(\mathbf{1},\mathbf{5})=\widetilde{A}_{0ij}(\mathbf{1},\mathbf{5})+g_{0}\sum_{k,l}\int d\mathbf{3}\widetilde{L}_{ikkj}(\mathbf{1},\mathbf{3})\widetilde{A}_{ll}(\mathbf{3},\mathbf{5})-g_{0}\sum_{k,l}\int d\mathbf{3}\widetilde{L}_{iklj}(\mathbf{1},\mathbf{3})\widetilde{A}_{kl}(\mathbf{3},\mathbf{5}),
  \label{eq:Main-11}
\end{equation}
\end{widetext}
where $\widetilde{L}_{iklj}(\mathbf{1},\mathbf{3})=G_{ik}(\mathbf{1},\mathbf{3})\tilde{G_{lj}}(\mathbf{3},\mathbf{1})$
and $\widetilde{A}_{0ij}(\mathbf{1},\mathbf{5})$ is correspondingly
expressed via $\widetilde{L}_{iklj}(\mathbf{1},\mathbf{3})$.

\subsection{Spherical symmetry and the angular momentum}
\label{sub:Symmetry-of-angular}

Solving even the mean-field density response in Eq.~\eqref{eq:Main-11}
in a spatially inhomogeneous system such as the trapped gas is a numerically
very demanding task. However, taking advantage of the assumed spherical
symmetry of the system, the Eq.~\eqref{eq:Main-11} can be greatly
simplified by expressing all quantities in spherical coordinates.
The spherical symmetry implies separation of different angular momentum
responses, allowing a significant simplification of the numerical
calculations.

Expressing the response function using Legendre polynomials 
(and using the knowledge that the response function depends only on the
time difference $t_1-t_2$) yields
\begin{equation}
  \tilde{A}_{ik}(\mathbf{1},\mathbf{2}) = \sum_{L,\omega}\mathcal{A}_{ik,L}(r_{1},r_{2},\omega)P_{L}(\cos\gamma)e^{-i\omega(t_{1}-t_{2})},\label{eq:a-2}\end{equation}
 where $\omega$ is a frequency (can be either real or imaginary),
$P_{L}(\cos\gamma)$ is the Legendre polynomial, and $\gamma$ is
the angle between the vectors $\mathbf{r}_{1}$ and $\mathbf{r}_{2}$.
Other functions involved in the solution, such as the Green's function
$\tilde{G}_{ik}(\mathbf{1},\mathbf{3})$ and $L_{iklj}(\mathbf{1},\mathbf{2},\mathbf{3})$
can be decomposed in the same way. Instead of eight parameters $\mathbf{r}_{1},t_{1},\mathbf{r}_{2},t_{2}$
all the functions now depend only on four parameters: $r_{1},r_{2}$
(magnitudes of the vectors $\mathbf{r}_{1}$ and $\mathbf{r}_{2}$),
$\omega$ (frequency) and $L$ (angular momentum).

With this decomposition, Eq.~\eqref{eq:Main} simplifies to
\begin{equation}
\begin{split}
  &\mathcal{A}_{ij,L}(r_{1},r_{5},\omega)=\mathcal{A}_{0ij,L}(r_{1},r_{5},\omega)\\
&+g_{0}\frac{4\pi}{2L+1}\int r_{3}^{2}dr_{3}\mathcal{L}_{ikkj,L}(r_{1},r_{3},\omega)\mathcal{A}_{ll,L}(r_{3},r_{5},\omega)\\
&-g_{0}\frac{4\pi}{2L+1}\int r_{3}^{2}dr_{3}\mathcal{L}_{iklj,L}(r_{1},r_{3},\omega)\mathcal{A}_{kl,L}(r_{3},r_{5},\omega).
  \label{eq:main_symmetry}
\end{split}
\end{equation}

As explained in Section~\ref{sub:Frequency-of-collective}, the peaks
in the density response $\mathcal{A}_{ij,L}(r_{1},r_{5},\omega)$
yield the frequencies of collective excitations, now corresponding
to a specific angular momentum $L$. For example, in this work we will
study the case $L=0$, in other words the spherically symmetric monopole
mode collective excitation.

In the Section~\ref{sub:Diagonalisation-of-Hamiltonian} we calculate
the coefficient $\mathcal{L}_{ikkj,L}(r_{1},r_{3},\omega)$ via the 
Bogoliubov-deGennes approximation. In Section~\ref{sub:Basic-definitions-for} 
we 
consider the
simplifications we need to proceed from the analytical equations to
the numerical calculations.

\subsection{Green's functions in the Bogoliubov-deGennes approach}
\label{sub:Diagonalisation-of-Hamiltonian}

In this section we will use the mean-field Bogoliubov-deGennes (BdG)
method to obtain the single particle Green's functions in the trap, thus allowing the
numerical solution of the density response. The Bogoliubov-deGennes
equations can be obtained by approximating the interaction part of
the Hamiltonian \eqref{eq:U-prev} by a quadratic form 
\begin{equation}
  U=-\int d\mathbf{r}~\psi_{\uparrow}^{\dagger}(\mathbf{r})\psi_{\downarrow}^{\dagger}(\mathbf{r})\Delta\left(\mathbf{r}\right)+H.c.,
  \label{eq:U_appr}
\end{equation}
 where $\Delta\left(\mathbf{r}\right)=\left|g_{0}\right|\left\langle \psi_{\uparrow}(\mathbf{r})\psi_{\downarrow}(\mathbf{r})\right\rangle $
is the order parameter. The Hartree energy is $g_{0}\int d\mathbf{r}\left(\psi_{\uparrow}^{\dagger}(\mathbf{r})\psi_{\uparrow}(\mathbf{r})n_{\downarrow}(\mathbf{r})+\psi_{\downarrow}^{\dagger}(\mathbf{r})\psi_{\downarrow}(\mathbf{r})n_{\uparrow}(\mathbf{r})\right)$,
where $n_{\uparrow\left(\downarrow\right)}(\mathbf{r})$ is the density
of up(down) particles. Most of the results below neglect 
the Hartree energy because, especially for strong interactions
$|k_\mathrm{F}a| \sim 1$, the inclusion of the Hartree energy
would require a prohibitively high cutoff energy, which is a critical
quantity for the efficiency of the numerical method. However,
we will briefly consider below the effect of the Hartree energy shift in the
weakly interacting regime.

The present work is concentrated in the moderate to weak interaction
regime $|k_\mr{F} a| \lesssim 1$, and the effect of the Hartree energy 
is to compress the gas slightly, increasing the order parameter at the
centre of the trap~\cite{Jensen2007a}. 
As will be seen in Section \ref{sec:results}, the key
quantity for the density response is the order parameter profile instead
of the interaction strength. Consequently the results will be shown
as a function of the order parameter at the centre of the trap $\Delta(r=0)$.
The compressing effect of the Hartree energy is thus relevant for
finding the precise correspondence between the system input parameters
(atom numbers, masses, trap frequencies, temperature, and interaction 
strength) and the order parameter. However, as will be seen, the Hartree
energy does not change the qualitative features of the density response. 

With these approximations the Hamiltonian in Eq.~\eqref{eq:Hamiltonian}
becomes quadratic in $\psi_{\alpha}(\mathbf{r})$: 
\begin{equation}
\begin{split}
  H=&\sum_{\alpha=\left\{ \uparrow,\downarrow\right\} }\int d\mathbf{r}\psi_{\alpha}^{\dagger}(\mathbf{r})\Bigl[-\frac{\hbar^2 \nabla^2}{2m}-\mu+\frac{m\omega_{\mr T}^{2}r^{2}}{2}\Bigr]\psi_{\alpha}(\mathbf{r})\\
&+\int d\mathbf{r}~\psi_{\uparrow}^{\dagger}(\mathbf{r})\psi_{\downarrow}^{\dagger}(\mathbf{r})\Delta\left(\mathbf{r}\right)+H.c.
  \label{eq:H_simpl}
\end{split}
\end{equation}
 The fermion fields can be expressed in the harmonic oscillator basis
as $\psi_{\alpha}(\mathbf{r})=\sum_{nlm}R_{nl}(r)Y_{lm}(\theta)c_{nlm\alpha}$,
where the operator $c_{nlm\sigma}$ destroys an atom from the harmonic
oscillator eigenstate $nlm$, $Y_{lm}(\theta)$ are the spherical
harmonics, and the radial eigenstates are given by 
\begin{equation}
  R_{nl}(r)=\sqrt{2}(m\omega_{\mr T})^{3/4}\sqrt{\frac{n!}{(n+l+1/2)!}}e^{-\bar{r}^{2}/2}\bar{r}^{l}L_{n}^{l+1/2}(\bar{r}^{2}),
\end{equation}
 where $L_{n}^{l+1/2}(\bar{r}^{2})$ is the associated Laguerre polynomial
and $\bar{r}\equiv r\sqrt{\frac{m\omega_{\mr T}}{\hbar}}$.

The Hamiltonian in Eq.~\eqref{eq:H_simpl} can be diagonalized using a canonical
transformation 
\begin{equation}
  c_{nlm\uparrow}=\sum_{j=1}^{N}W_{n,j}^{l}\gamma_{jlm\uparrow}+(-1)^{m}\sum_{j=1}^{N}W_{n,N+j}^{l}\gamma_{jl-m\downarrow}^{\dagger}
  \label{eq:c_nlm}
\end{equation}
and
\begin{equation}
  c_{nlm\downarrow}^{\dagger}=(-1)^{m}\sum_{j=1}^{N}W_{N+n,j}^{l}\gamma_{jl-m\uparrow}+\sum_{j=1}^{N}W_{N+n,N+j}^{l}\gamma_{jlm\downarrow}^{\dagger}
  \label{eq:c^dag_nlm}
\end{equation}
which yields the diagonalized Hamiltonian 
\begin{equation}
  H=\sum_{jlm,\alpha}E_{jl}\gamma_{jlm\alpha}^{\dagger}\gamma_{jlm\alpha}.
  \label{eq:H_}
\end{equation}

The index $j$ in $\gamma_{jlm\alpha}$ corresponds to the enumeration
of the quasiparticle states, the index $l$ is the angular momentum,
and $m=-l,-l+1,\ldots, l$ is the $z$-component of the angular momentum.
The index $\alpha$ does not have the meaning of a physical (pseudo)spin 
anymore, but it has an auxiliary function.

\begin{figure}
  \centering
  \includegraphics[width=0.45\textwidth]{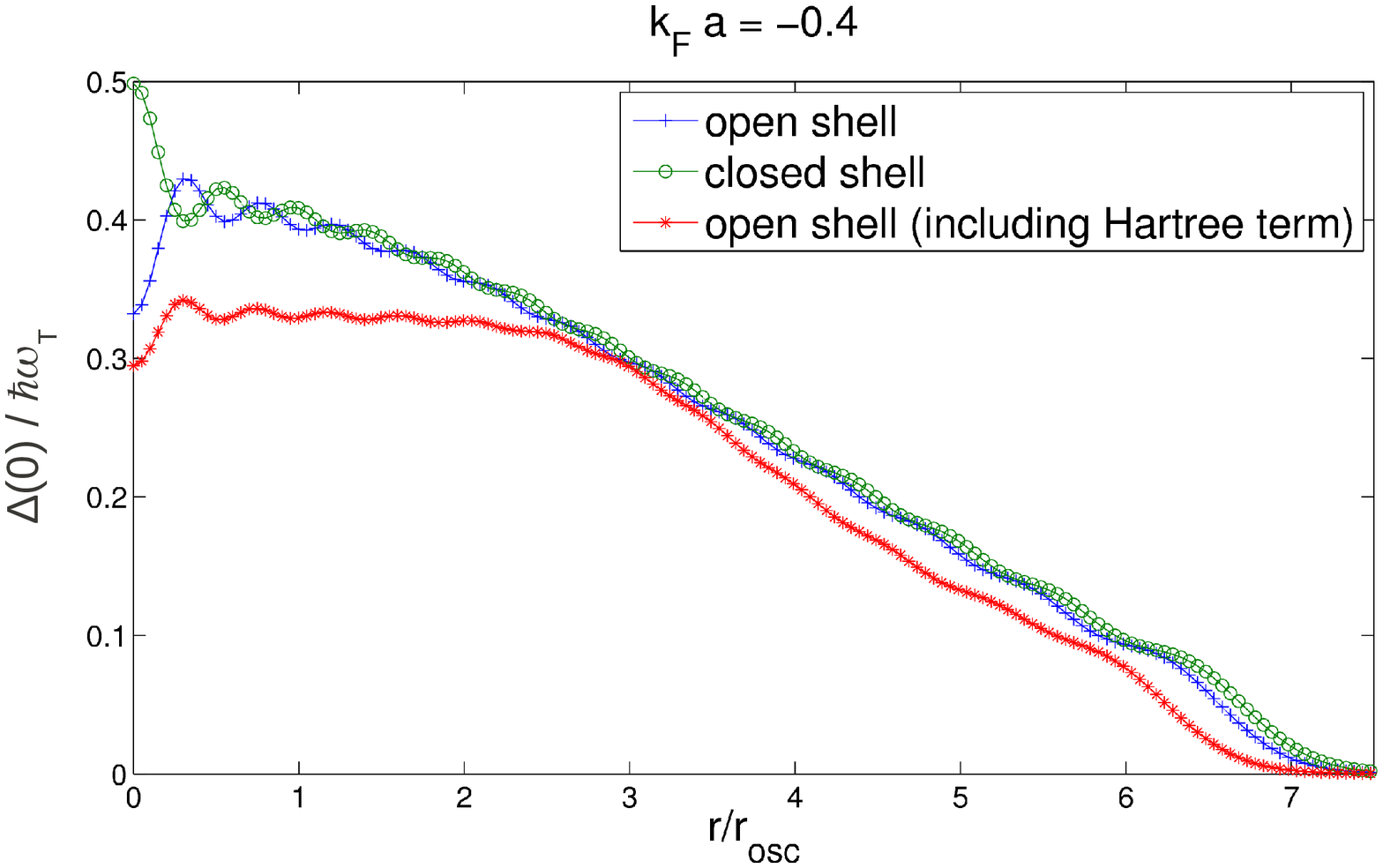}
  \includegraphics[width=0.45\textwidth]{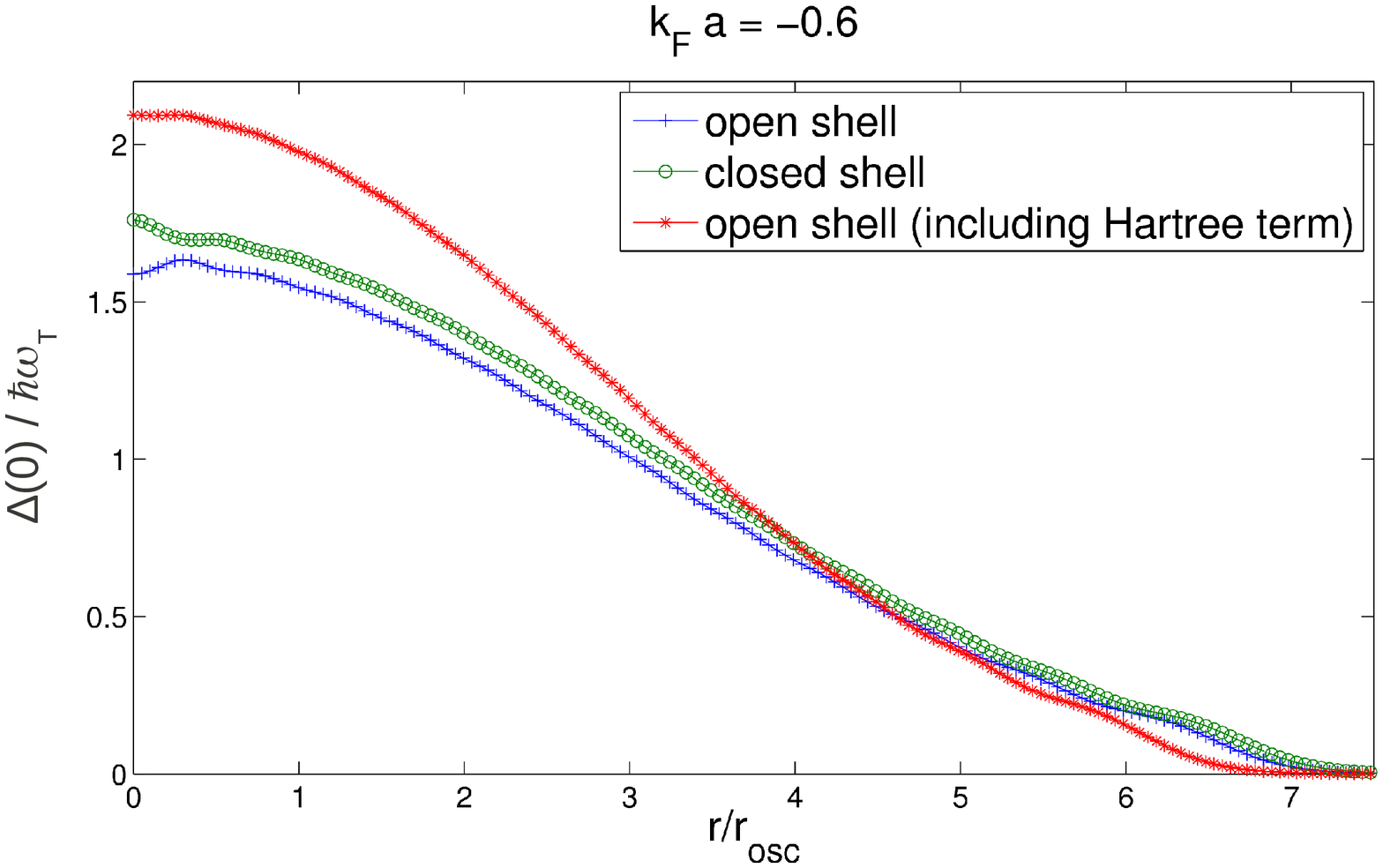}
  \includegraphics[width=0.45\textwidth]{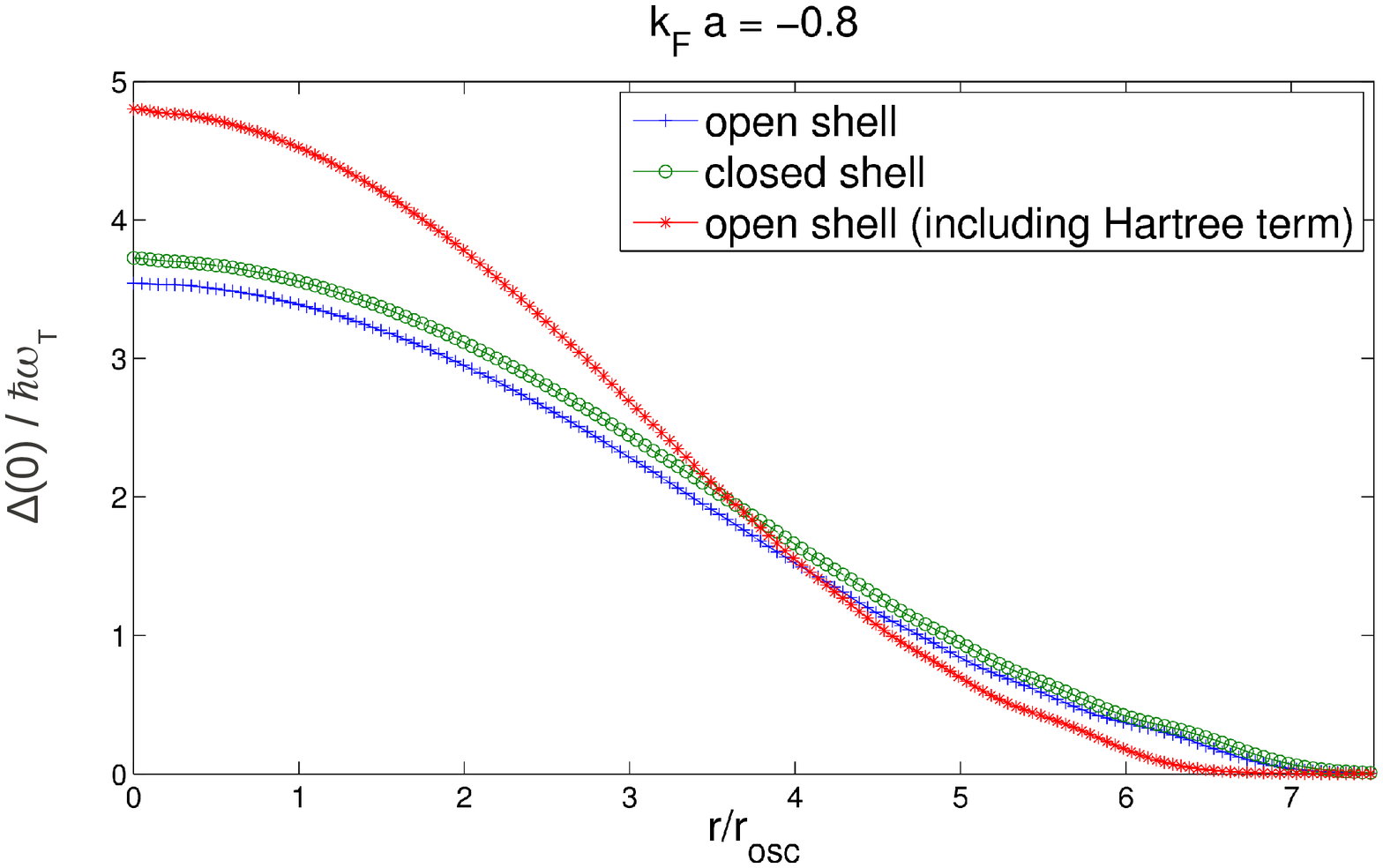}
  \caption{(Color online) Order parameter profiles $\Delta(r)$ obtained using the
Bogoliubov-deGennes method for different interaction strengths. 
Open and closed shell data correspond to different total atom numbers, 
$N = 4930$ and $N = 5600$, respectively. Shown are 
also the profiles that include the Hartree energy shift contribution, 
calculated for $N=4930$. Figures show data for $k_F a = -0.4$, $k_F a = -0.6$ and $k_F a = -0.8$.}
  \label{fig:0}
\end{figure}

The equation for the order parameter profile at zero temperature is 
\begin{equation}
 \Delta(r) = g_0 \sum_{nn'lj} \frac{2l+1}{4\pi} R_{nl} (r) R_{n'l} (r) W_{n,j}^l W_{N+n',j}^l,
\end{equation}
which needs to be solved self-consistently together with the number
equations for the local atom densities
\begin{equation}
  n_\uparrow(r) = \sum_{nn'lj} \frac{2l+1}{4\pi} R_{nl} (r) R_{n'l} (r) W_{n,j}^l W_{n',j}^l,
\end{equation}
and
\begin{equation}
  n_\downarrow(r) = \sum_{nn'lj} \frac{2l+1}{4\pi} R_{nl} (r) R_{n'l} (r) W_{n+N,j}^l W_{n'+N,j}^l.
\end{equation}
Fig.~\ref{fig:0} shows the order parameter profiles $\Delta(r)$ 
for different atom numbers, interaction strengths, and also with the 
Hartree energy shift.

We calculate the coefficients $W_{n,j}^{l}$ and the eigenenergies
$E_{jl}$ numerically. The Green's function $G_{ij}(\mathbf{1},\mathbf{2})$
can be expressed via $W_{n,j}^{l}$:
\begin{equation}
\begin{split}
  \widehat{G}&(\mathbf{r}_{1},\mathbf{r}_{2},\Omega_{n})=-\sum_{l}\frac{2l+1}{4\pi}P_{l}(\cos\theta_{12}) \times \\
&\times \left( \frac{\Lambda_{jl}^{-}(r_{1})\Lambda_{jl}^{-\dagger}(r_{2})}{i\Omega_{n}-E_{jl}}+\frac{\Lambda_{jl}^{+}(r_{1})\Lambda_{jl}^{+\dagger}(r_{2})}{i\Omega_{n}+E_{jl}}\right),
  \label{eq:a-3}
\end{split}
\end{equation}
where $\Lambda_{jl}^{-}(r)=\sum_{n}\left(\begin{array}{c}
W_{n,N+j}^{l}\\
W_{N+n,N+j}^{l}\end{array}\right)R_{nl}(r)$, $\Lambda_{jl}^{+}(r)=\sum_{n}\left(\begin{array}{c}
W_{n,j}^{l}\\
W_{N+n,j}^{l}\end{array}\right)R_{nl}(r)$, and $P_{l}(\cos\theta_{12})=\frac{4\pi}{2L+1}\sum_{M=-L}^{L}Y_{LM}^{*}(\theta_{1},\varphi_{1})Y_{LM}(\theta_{2},\varphi_{2})$
are the Legendre polynomials and $\theta_{12}$ is the angle between
the vectors $\mathbf{r}_{1}$ and $\mathbf{r}_{2}$. Here $\widehat{G}$
is the Green's function in the matrix form $\widehat{G}=\left(\begin{array}{cc}
G_{\uparrow\uparrow} & G_{\uparrow\downarrow}\\
G_{\downarrow\uparrow} & G_{\downarrow\downarrow}\end{array}\right)$. 

The kernel $\mathcal{L}_{iklj,L}(r_{1},r_{3},\omega)$ of Eq.~\eqref{eq:main_symmetry} is (see Appendix~\ref{sec:Coefficients--via}):
\begin{widetext}
\begin{equation}
\begin{split}
\mathcal{L}_{iklj,L}(r_{1},r_{3},\omega)=&\left(2L+1\right)\sum_{L_{1}L_{2}}\left(\begin{array}{ccc} 
     L & L_{1} & L_{2}\\
     0 & 0 & 0 
  \end{array}\right)^{2} \frac{2L_{1}+1}{4\pi}\frac{2L_{2}+1}{4\pi} \\
&*\sum_{J_{1}J_{2}}\left(\lambda_{J_{1}L_{1},ik}^{-}\lambda_{J_{2}L_{2},lj}^{-}\frac{n_{\mr F}(E_{J_{1}L_{1}})-n_{\mr F}(E_{J_{2}L_{2}})}{\omega+E_{J_{1}L_{1}}-E_{J_{2}L_{2}}}+\lambda_{J_{1}L_{1},ik}^{+}\lambda_{J_{2}L_{2},lj}^{+}\frac{n_{\mr F}(-E_{J_{1}L_{1}})-n_{\mr F}(-E_{J_{2}L_{2}})}{\omega-E_{J_{1}L_{1}}+E_{J_{2}L_{2}}}\right.\\
&+\left.\lambda_{J_{1}L_{1},ik}^{-}\lambda_{J_{2}L_{2},lj}^{+}\frac{n_{\mr F}(E_{J_{1}L_{1}})-n_{\mr F}(-E_{J_{2}L_{2}})}{\omega+E_{J_{1}L_{1}}+E_{J_{2}L_{2}}}+\lambda_{J_{1}L_{1},ik}^{+}\lambda_{J_{2}L_{2},lj}^{-}\frac{n_{\mr F}(-E_{J_{1}L_{1}})-n_{\mr F}(E_{J_{2}L_{2}})}{\omega-E_{J_{1}L_{1}}-E_{J_{2}L_{2}}}\right).
\label{eq:a-4}
\end{split}
\end{equation}
\end{widetext}
 Here the occupation numbers are given by the Fermi-Dirac equation
$n_{\mr F}(E)=\frac{1}{\exp(\beta E)+1}$ at the temperature $k_{B}T=\frac{1}{\beta}$.
Furthermore, $\left(\begin{array}{ccc}
L & L_{1} & L_{2}\\
0 & 0 & 0\end{array}\right)$ are the Wigner 3j-symbols. Finally, 
$\lambda_{J_{1}L_{1},ik}^{\pm}=\Lambda_{J_{1}L_{1},i}^{\pm}(r_{1})\Lambda_{J_{1}L_{1},k}^{\pm\dagger}(r_{3})$
and $\lambda_{J_{2}L_{2},lj}^{\pm}=\Lambda_{J_{2}L_{2},l}^{\pm}(r_{3})\Lambda_{J_{2}L_{2},j}^{\pm\dagger}(r_{1})$.

In this section we have derived the kernel of the Eq.~\eqref{eq:main_symmetry},
$\mathcal{L}_{iklj,L}(r_{1},r_{3},\omega)$, as a function of the
coefficients $W_{n,j}^{l}$ and the eigenenergies $E_{jl}$, which
are to be calculated numerically. Thus Eq.~\eqref{eq:main_symmetry}
can be solved, and the resulting density response can be obtained.

\subsection{Basic definitions for numerical calculations}
\label{sub:Basic-definitions-for}

In this section we will discuss the parameters used in the numerical
calculations. As noted in Eq.~\eqref{eq:K}, the gas of atoms of mass $m$
is confined in the harmonic trapping potential
of a frequency $\omega_{\mr T}$. The system is characterised by two units:
the unit of energy (the difference between neighbouring levels) $\hbar\omega_{\mr T}$
and the unit of length (the oscillator length) $r_\mr{osc}=$$\sqrt{\frac{\hbar}{m\omega_{\mr T}}}$.
In the numerical calculations we use dimensionless values, in the
unit system based on $\hbar\omega_{\mr T}$ and $r_\mr{osc}$.

The interaction strength is determined by the two-body scattering length
$a$ as $1/g_0 = \frac{m}{4\pi\hbar^2 a}  - \gamma(r)$
where the position dependent renormalization coefficient is calculated
in the local density approximation~\cite{Grasso2003a} and is given by
\begin{equation}
  \gamma(r)=\frac{mk_\mr{c}(r)}{2\pi^2\hbar^2} \left( \frac{\kappa(r)}{2} \log \frac{1+\kappa(r)}{1-\kappa(r)} -1 \right)
\end{equation}
where $\frac{\hbar^2 k_\mr{c}(r)^2}{2m} = E_\mr{c}-\frac{1}{2}m\omega_\mr{T}^2 r^2$, $\frac{\hbar^2 k_\mr{F}(r)^2}{2m} = \mu-\frac{1}{2} m\omega_\mr{T}^2 r^2$, $\kappa (r)=k_\mr{F}(r)/k_\mr{c}(r)$, and
$E_\mr{c}=\hbar \omega_\mr{T}\left(2N_\mr{c}+1\right)$ is the cut-off energy. 
We use cut-off energy $E_\mr{c}=60\,\hbar\omega_{\mr T}$ (for runs with the Hartree energy we used $E_\mr{c}=100\,\hbar\omega_{\mr T}$), which proved to be
sufficient for studying the density response of the rather weakly interacting gas,
resulting in at most $1\,\%$ error in the magnitude of the pairing field. 
The Fermi energy is $E_\mr{F}=24\,\hbar\omega_{\mr T}$ for $N=4930$. 
For the interaction in the density response in Eq.~\eqref{eq:main_symmetry}
we use the above renormalized interaction $g_0$ as in Ref.~\cite{Grasso2005a}.

\section{Results}
\label{sec:results}

In this section we present the numerical results of the density response.
In Section~\ref{sub:What-do-we} we relate the density response with
the frequency of collective excitations. In Section~\ref{sub:Results-for-weak}
we study the density response function for weak interactions, and in 
Section~\ref{sub:damping} we interpret the width of a band of collective
excitations as the damping rate of the modes. In the
Section~\ref{sub:Interaction-dependence} we show how the frequencies
of the excitations depend on the interaction strength and discuss the interesting
effect of merging of the pair vibration and collisionless hydrodynamic excitation branches.
In the Section~\ref{sub:Open-and-closed} we discuss the cases of 
closed and open shells (i.e. cases of different numbers of atoms).

\subsection{Spectrum of the monopole mode}
\label{sub:What-do-we}

Our goal is to study the density response $\mathcal{A}_{ij,L}(r_{1},r_{5},\omega)$,
which was defined in Section~\ref{sub:Symmetry-of-angular}. As
was discussed in Section~\ref{sub:Frequency-of-collective}, the
peaks in the response function yield the frequencies of the collective
excitations of the system. In order to better understand the physical origin of various excitations
we will consider also the non-RPA response $\mathcal{A}_{0ij,L}(r_{1},r_{5},\omega)$,
which was discussed in Section~\ref{sub:Symmetry-of-angular}.
The peaks in this function reflect the frequencies of single particle
excitations. Thus in this article we will call it the single particle
density response, and $\mathcal{A}_{ij,L}(r_{1},r_{5},\omega)$ the
full density response. The simultaneous study of $\mathcal{A}_{ij,L}(r_{1},r_{5},\omega)$
and $\mathcal{A}_{0ij,L}(r_{1},r_{5},\omega)$ allows one to identify
the origin of different collective excitations; some excitations in
$\mathcal{A}_{ij,L}(r_{1},r_{5},\omega)$ have their origin in $\mathcal{A}_{0ij,L}(r_{1},r_{5},\omega)$
as corresponding single particle excitations, while others, which we refer
to as purely collective excitations, do not have a corresponding peak
in $\mathcal{A}_{0ij,L}(r_{1},r_{5},\omega)$.

The density response $\mathcal{A}_{ij,L}(r_{1},r_{5},\omega)$ depends
on six parameters: spin indices $i$ and $j$, the angular momentum
$L$, positions $r_{1}$ and $r_{5}$, and the frequency $\omega$.
In addition, the response can be calculated for three different kinds
of probing fields $\phi_{\uparrow}$, $\phi_{\downarrow}$ and $\eta$,
as described in Eq.~\eqref{eq:Perturbation}. However, in Section~\ref{sub:Spectrum-as-function} 
we will show that the main features of the collective excitation spectra
do not depend on the parameters $i$, $j$,
$r_{1}$, $r_{5}$ nor on the fields $\phi_{\uparrow}$, $\phi_{\downarrow}$,
$\eta$, but only on the angular momentum $L$ and the frequency
$\omega$.
In this paper we will study only the monopole mode, corresponding
to $L=0$. Nevertheless, our method allows also the calculation of
the response function for higher angular momenta $L>0$. Hence we
will now limit the study to $\left.\mathcal{A}_{\uparrow\uparrow}(r_1=0,r_5=0,\omega)\right|_{\phi_{\uparrow},L=0}$, denoting it as $\mathcal{A}(\omega)$.

\begin{figure}
  \includegraphics[width=0.45\textwidth]{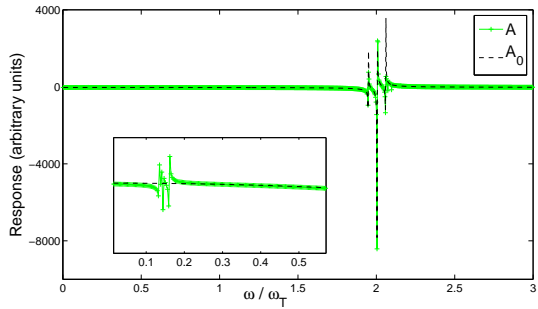}
  \caption{(Color online) The density response in the extreme weakly interacting limit $k_{\mr F}a=-0.22$ ($\Delta\left(0\right)=0.06\,\hbar\omega_{\mr T}$)
shows two groups of excitations. One, a prominent one, at frequencies $\omega \approx 2\,\omega_{\mr T}$ and another (shown enlarged
in the inset) at $\omega \approx 0.15\,\omega_{\mr T}$. Shown are the full density response $\mathcal{A} (\omega)$ and the single particle
response $\mathcal{A}_0(\omega)$.}  
  \label{fig:1}
\end{figure}

\begin{figure}
  \includegraphics[width=0.45\textwidth]{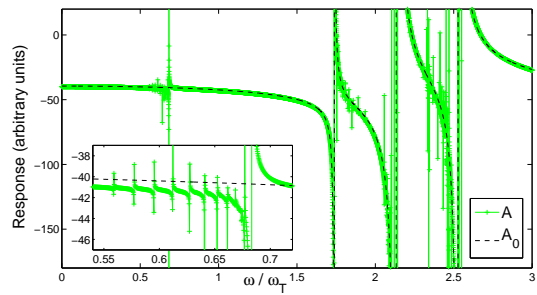}
  \caption{(Color online) The density response for a slightly stronger interaction than in Fig.~\ref{fig:1} $k_{\mr F}a=-0.50$ ($\Delta\left(0\right)=0.78\,\hbar\omega_{\mr T}$), shows the increase in the widths of the collective excitation bands (in $\mathcal{A} (\omega)$) 
and the separation of the single particle excitations (in $\mathcal{A}_0 (\omega)$).}
  \label{fig:2}
\end{figure}

\subsection{Results for weak interactions}
\label{sub:Results-for-weak}

In this chapter we study the full density response 
$\mathcal{A}_{\uparrow\uparrow}(\omega)=\frac{\delta\rho_{\uparrow}}{\delta\phi_{\uparrow}}$
and the corresponding single particle density response $\mathcal{A}_{0,\uparrow\uparrow}(\omega)$,
which we denote as $\mathcal{A}(\omega)$ and $\mathcal{A}_{0}(\omega)$,
respectively. From here on until further notice we consider the response
for the total number of atoms $N=4930$.

Typical results for the density response in the weakly interacting
regime are shown in Figs.~\ref{fig:1} and~\ref{fig:2}. Fig.~\ref{fig:1} 
shows the full density response $\mathcal{A}(\omega)$
and the corresponding single particle density response $\mathcal{A}_{0}(\omega)$
for $k_{\mr F}a=-0.22$. This case corresponds to the gap in the centre
of the trap $\Delta\left(0\right)=0.06\,\hbar\omega_{\mr T}$. In Fig.~\ref{fig:2} 
are shown responses for a higher interaction $k_{\mr F}a=-0.50$
(the gap in the centre of the trap $\Delta\left(0\right)=0.78\,\hbar\omega_{\mr T}$).

Both figures show two groups of peaks, one at low frequencies $\omega < 2\,\omega_{\mr T}$
and another close to $2\,\omega_{\mr T}$. This is a typical result for sufficiently
weak interactions, with the key criterion being the value of the gap 
$\Delta (0) < 2\,\hbar\omega_{\mr T}$. The group of peaks in the vicinity of $2\,\omega_{\mr T}$
is present both in the full response $\mathcal{A}(\omega)$ as well as in the
single particle response $\mathcal{A}_{0} (\omega)$, the corresponding single particle 
excitations describing transitions of single atoms from the Fermi surface to the next
higher oscillator energy level. These excitations are described by the hydrodynamic
model for a collisionless gas.
The other group of peaks at lower frequencies is 
shown in the insets of Figs.~\ref{fig:1} and~\ref{fig:2}. These peaks are present
only in the full density response $\mathcal{A}(\omega)$ but not in the single particle
density response $\mathcal{A}_{0}(\omega)$ and thus these are purely collective
excitations. These are called pair vibration modes and they describe pair amplitude
modulations~\cite{Bruun2002a}. In the very weakly interacting limit the pair vibration
mode frequency is given by the pairing gap as $\omega \approx 2\Delta (0)/\hbar$, but
as already seen in Fig.~\ref{fig:2} the relation does not hold for stronger
interactions (see the positions of the $\mathcal{A}(\omega)$ peaks in Figs.~\ref{fig:1}
and~\ref{fig:2} compared to $2\Delta(0)$).

Note that for the pair vibration modes we do not observe
an individual peak but rather a group of peaks close to each other. With
the interaction increasing, the peaks are shifted to higher frequencies
and the distance between them increases. We discuss the interaction dependence
more in Section~\ref{sub:Interaction-dependence}.

Also the group of peaks at $\omega \approx 2\,\omega_{\mr T}$ experiences significant 
changes when the interaction is increased. Fig.~\ref{fig:2} shows
how for stronger interactions the three single particle excitations 
are accompanied by at least $17$ peaks, corresponding to collective
modes, between $\omega=2.01\,\omega_{\mr T}$ and $\omega=2.78\,\omega_{\mr T}$.

\subsection{Damping rate}
\label{sub:damping}

\begin{figure}
  \includegraphics[width=0.45\textwidth]{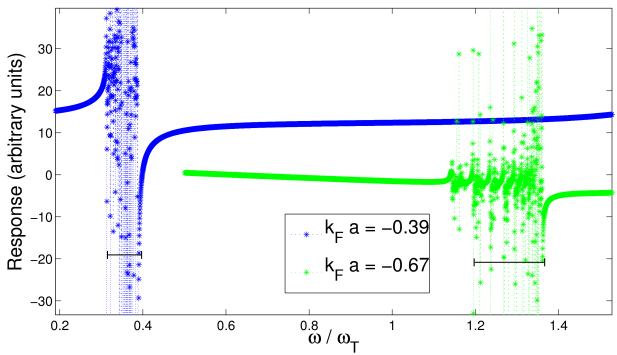}
  \caption{(Color online) The pair vibration modes, corresponding to purely collective 
excitations, for $k_{\mr F}a=-0.39$ ($\Delta\left(0\right)=0.30\,\hbar\omega_{\mr T}$) and $k_{\mr F}a=-0.67$ ($\Delta\left(0\right)=2.26\,\hbar\omega_{\mr T}$). Shown are also 
the corresponding collective mode band widths $0.06\,\omega_{\mr T}$ and $0.22\,\omega_{\mr T}$, respectively.}
  \label{fig:comparison}
\end{figure}

Excitations with similar frequencies can be coupled, resulting in a
damping of the modes. Even though the excitations for a finite system manifest
as poles with real energies and vanishing imaginary parts, in practice
a large number of collective excitations with nearby lying frequencies
cannot be distinguished.
The distance between the mode frequencies within the
band yields the time required for resolving the various peaks, but for
time scales shorter than this the resulting real time evolution 
appears damped. For short time scales, the characteristic damping 
rate is given by the width of the collective excitation band.
Fig.~\ref{fig:comparison} shows the pair
vibration modes for stronger interactions, revealing the gradual increase
in the collective excitation band width and thus the increase in the
corresponding damping rate. For $k_{\mr F}a = -0.39$ ($\Delta(0) = 0.3\,\hbar\omega_{\mr T}$)
the width is roughly $0.06\,\omega_{\mr T}$ and for $k_{\mr F}a = -0.67$ ($\Delta(0) = 2.26\,\hbar\omega_{\mr T}$)
the width is $0.22\,\omega_{\mr T}$.

\begin{figure}
  \includegraphics[width=0.45\textwidth]{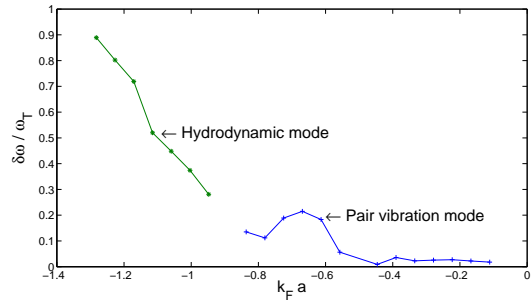}
  \caption{(Color online) The lowest frequency collective excitation band width as
a function of interactions. The band width increases rapidly when the
order parameter $\Delta(0)$ approaches $4\,\hbar \omega_\mr{T}$ at
$k_\mr{F}a \approx -0.8$.}
  \label{fig:damping}
\end{figure}

Fig.~\ref{fig:damping} shows the width of the lowest frequency collective
excitation band (the pair vibration mode for weak interactions) as a 
function of interactions (both as a function of $k_{\mr F}a$ and $\Delta(0)$). 
While for weak interactions the band is narrow, the width increases rapidly 
when approaching the crossover regime $k_\mr{F}a \approx -0.8$, 
corresponding to $\Delta(0) \approx 4\,\hbar \omega_{\mr T}$. Close to 
the crossover, where the pair vibration and the hydrodynamic modes merge, 
the width of only the lowest excitation band may not be a good measure of the
damping as the distance between the two collective mode bands is less
than the widths of the two bands.

The number of peaks within the collective mode band depends on the system 
size, with larger systems yielding more peaks (the procedure of defining ``a peak'' 
is described in Section~\ref{sub:Procedure-of-defining}). In the thermodynamic limit 
$N \rightarrow \infty$, we expect that the collective mode band no longer consists of 
separate peaks but becomes continuous. However, as 
long as the magnitude of the pairing gap is unchanged the width of the band is 
unaffected. This conjecture is supported by our calculations with higher atom numbers (up to $N = 49300$).

\subsection{Interaction dependence}
\label{sub:Interaction-dependence}

\begin{figure*}
  \includegraphics[width=0.9\textwidth]{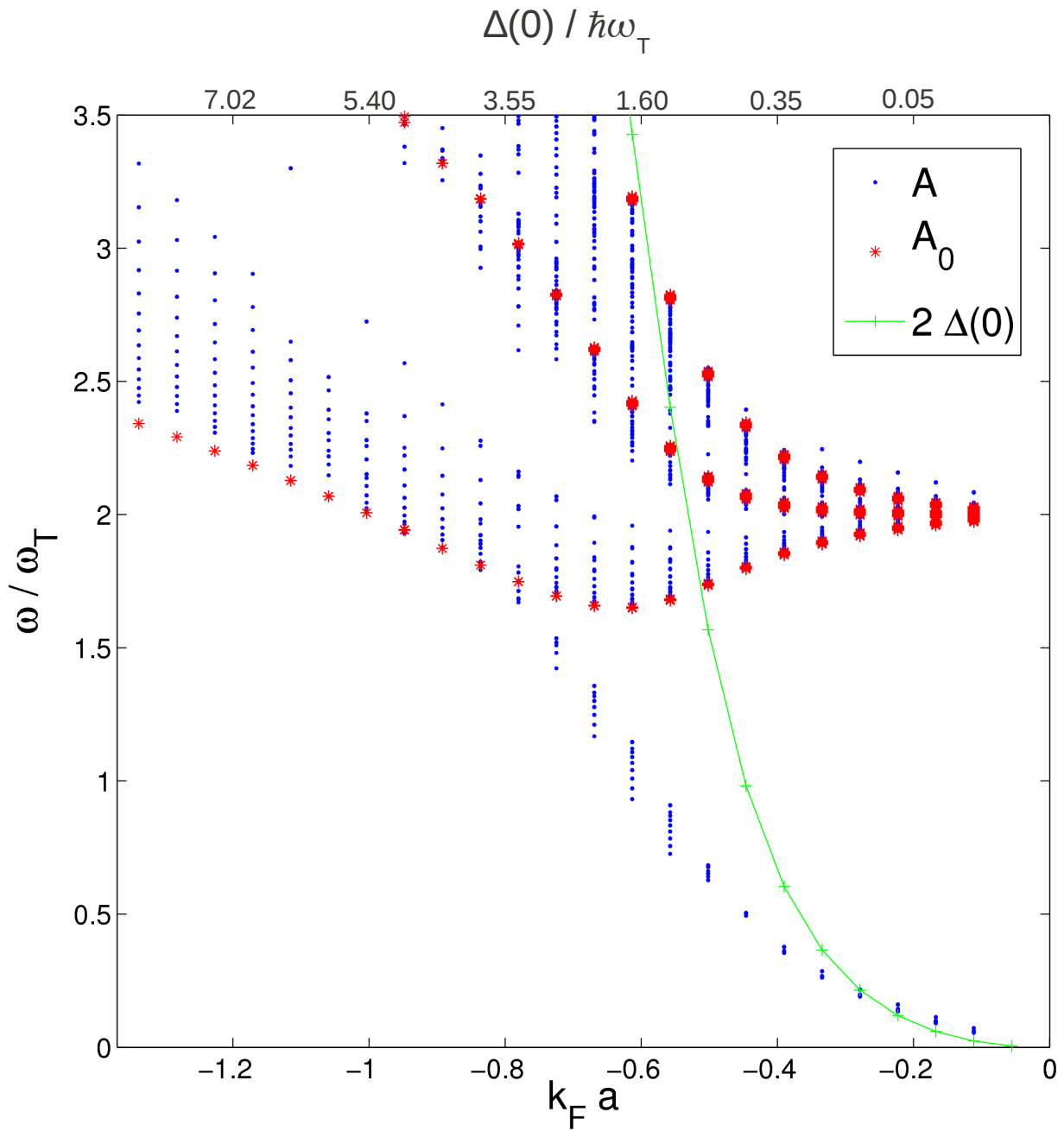}
  \caption{(Color online) The peaks in the density responses as a function of the interaction $k_{\mr F}a$ or $\Delta(0)$. The crossover from the pair vibration mode to the
strongly interacting hydrodynamic mode occurs for $\Delta(0) \approx 4\,\hbar\omega_{\mr T}$ when the low frequency pair vibration mode merges with the 
weakly interacting collisionless hydrodynamic mode.}
  \label{fig:closed_shell} 
\end{figure*}

When the interaction strength increases, the two branches, the pair vibration
modes and the collisionless hydrodynamic modes,
approach each other and eventually merge. Instead of the Fermi energy, the relevant
energy scale in this merging regime is the trap oscillator energy $\hbar \omega_\mr{T}$.
Subsequently the interaction strength is best measured as $\Delta(0)/\hbar \omega_\mr{T}$
instead of the standard $k_\mr{F}a$. 

The main result of this work, Fig.~\ref{fig:closed_shell}, shows how the 
positions of the peaks evolve
as a function of the interaction (showing both $k_{\mr F}a$ and $\Delta(0)$).
As discussed above, in the weakly interacting regime (here defined as the regime where
$\Delta(0) < 2\,\hbar \omega_{\mr T}$) there are two main branches of excitations.
One branch originates from the collisionless hydrodynamic excitation (or single particle
excitations) at frequency $\omega = 2\,\omega_{\mr T}$ and the other branch describes pair vibration modes, 
starting from the zero frequency. For weak interactions, the latter follows
the frequency $2\Delta(0)/\hbar$~\cite{Bruun2002a}, but the branch deviates
from this asymptotic value already at $\Delta(0) \approx 0.2\,\hbar\omega_{\mr T}$. The collisionless hydrodynamic and the pair vibration branches merge at 
around $k_\mathrm{F}a \approx -0.8$, corresponding to 
$\Delta(0) \approx 4\,\hbar\omega_{\mr T}$. For interactions
beyond that, the hydrodynamic and the pair vibration modes 
cannot be distinguished anymore and the two modes transform into the 
strongly interacting Goldstone mode in the strongly interacting limit.

The crossover between the pair vibration mode and the Goldstone mode takes 
place when the order parameter $\Delta(0)$ is of the order of a few trap 
oscillator energies $\hbar \omega_{\mr T}$. Since the key energy scale is given
by the trap frequency instead of the Fermi energy, the crossover is
realized at different interaction strengths $k_{\mr F}a$ if $\omega_{\mr T}$ 
is different. This was discussed already in Refs.~\cite{Bruun2002a,Grasso2005a}. The interaction strength studied in Ref.~\cite{Bruun2002a}
corresponded to $\Delta(0) < 2\,\hbar\omega_{\mr T}$, whereas 
in Ref.~\cite{Grasso2005a} the monopole mode was studied for $\Delta(0) \approx 6\,\hbar \omega_{\mr T}$. Here we consider the crossover region between 
these two limits.

Notice the depression of the collective mode frequencies in the crossover 
regime in Fig.~\ref{fig:closed_shell} and the increase of the collective 
mode band widths in Fig.~\ref{fig:damping}. These effects, the suppression 
of the collective mode frequency and the increase in the damping rate in 
the crossover regime where $\Delta(0) \sim \hbar\omega_{\mr T}$, are in 
qualitative agreement with similar effects observed in 
experiments~\cite{Bartenstein2004a,Kinast_hydrodynamics,Altmeyer2007b}
although the symmetries of both the trap and the perturbations are different 
in the experimental setups. The experiment in Ref.~\cite{Bartenstein2004a}
observed a dramatic increase in the damping rate and a decrease in the
radial compression mode frequency at a magnetic field $B \sim 910\,{\mr G}$
corresponding to the value $k_{\mr F}a \approx -0.5$. The radial breathing mode
was studied in Ref.~\cite{Kinast_hydrodynamics}, revealing a strong
damping rate maximum and a depression in the collective mode frequency
at a magnetic field $B \sim 1080\,{\mr G}$, corresponding to $k_{\mr F}a \approx -0.74$. Finally, the radial quadrupole mode was studied in Ref.~\cite{Altmeyer2007b}, exhibiting a damping rate maximum and a depression and a jump of the 
collective mode frequency at a magnetic field $B \approx 950\,{\mr G}$,
corresponding to $k_{\mr F}a \approx -0.8$. Despite the differences between
the experiments in the critical interaction strengths for observing the damping rate maxima, all three experiments are in the regime where the predicted
BCS pairing gap $\Delta(0)/\hbar$ is of the order of a few oscillator 
frequencies. However, a detailed comparison with the experiments would 
require a 
theoretical study of the same radial modes, not accessible in the present
model assuming the spherical symmetry.

\begin{figure*}
  \includegraphics[width=0.8\textwidth]{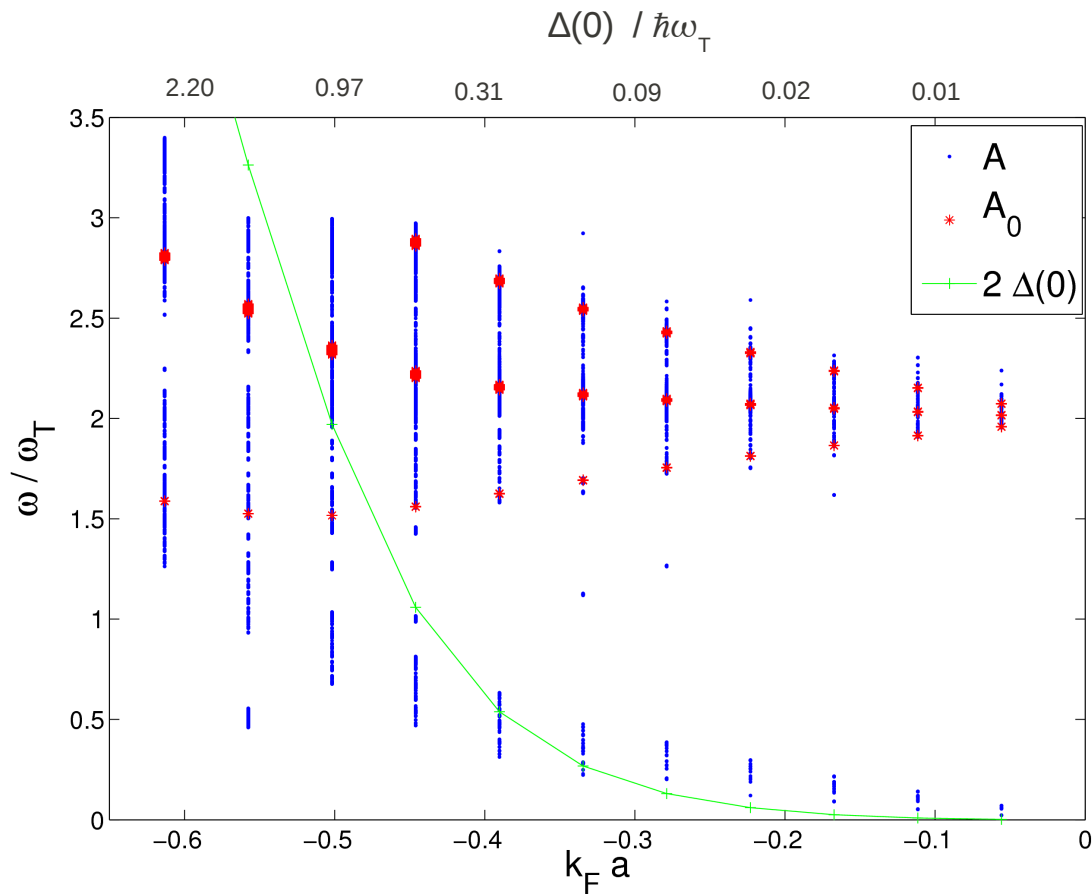}
  \caption{(Color online) The peaks in the density responses as a function of the interaction $k_{\mr F}a$ or $\Delta(0)$, including the Hartree energy
contribution. The Hartree energy compresses the gas, resulting in a larger
order parameter $\Delta(0)$. Comparing with Fig.~\ref{fig:closed_shell}, this shifts the crossover from the pair 
vibration mode to the strongly interacting hydrodynamic mode to
lower interaction strengths.}
  \label{fig:hartree_spectra} 
\end{figure*}

Fig.~\ref{fig:hartree_spectra} shows the effect of the Hartree energy
on the density response. The lowest energy collective mode band is shifted
to higher frequencies and it is wider. This results in the crossover between the two collective mode branches
occuring at a weaker interaction than when the Hartree energy was neglected. However, the qualitative features are 
unchanged from Fig.~\ref{fig:closed_shell}.

\subsection{Open and closed shells}
\label{sub:Open-and-closed}

Depending on the number of the atoms, the single particle density
response $\mathcal{A}_0$ can show significantly different behaviour. 
In this section we discuss the cases of closed and open 
shells~\cite{Bruun2002a}. The closed shell, which was the 
case discussed above with $N = 4930$,
corresponds to a case in which the uppermost occupied energy level
at the Fermi surface is fully occupied in the ideal noninteracting system.
In contrast, in an open shell configuration (for example $N=5600$ used below) 
only part of the energy level at the Fermi surface is occupied.

\begin{figure*}
  \includegraphics[width=0.9\textwidth]{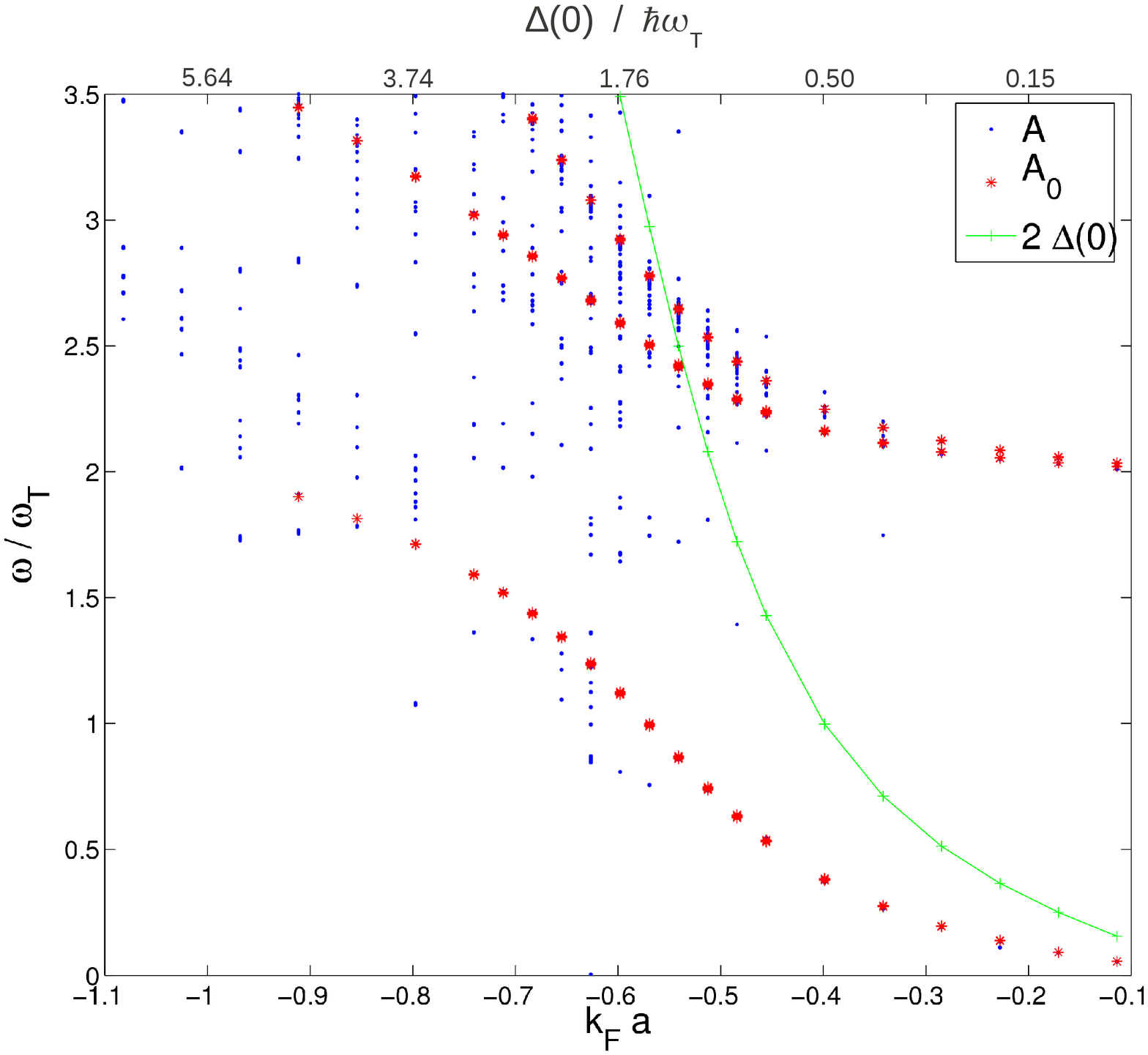}
  \caption{(Color online) The peaks of the density response as function of interactions
for open shell configuration. The full density response is similar
(albeit more noisy) to the closed shell configuration in Fig.~\ref{fig:closed_shell}, but here one of the single-particle excitation branches follows the pair
vibration mode frequency at weak interactions ($\Delta(0) < 4\,\hbar \omega_\mr{T}$).}
  \label{fig:open_shell}
\end{figure*}

The two cases give very different single particle density response functions
$\mathcal{A}_0$ in the weakly interacting limit ($\Delta(0) < 2\,\hbar \omega_\mr{T}$).
In the case of the closed shell, the lowest frequency single particle transition 
is the transition from the Fermi surface to the next
higher $n$-level, corresponding to the frequency $2\,\omega_{\mr T}$.
This results in a branch of single particle excitations starting at frequency
$2\,\omega_\mr{T}$ in Fig.~\ref{fig:closed_shell}. In the case of the open shell, the
energy level at the Fermi surface is also available for transitions, and the
lowest energy single particle transition is a pair breaking transition in which
the $n$ quantum number is not affected. The associated energy change 
is proportional to $2\,\Delta(0)$ in the weakly interacting limit, tending to zero
for vanishing interaction strength. The evolution of the peak positions as
a function of interactions for the open shell configuration is shown in the
Fig.~\ref{fig:open_shell}, revealing one single particle excitation branch
starting from zero frequency and two single particle branches from the frequency
$2\,\omega_\mr{T}$. While the difference between the two shell configurations
is a mesoscopic effect, the effect on the single particle excitation spectrum
is artificial as the experimentally observable quantity is not $\mathcal{A}_0$
but the full response $\mathcal{A}$. Indeed, comparing the full density responses
of the two configurations does not show any apparent difference, albeit the
data for the open shell configuration is somewhat more noisy due to numerical 
issues. 

\subsection{Other response functions}
\label{sub:Spectrum-as-function}

\begin{figure}
  \includegraphics[width=0.45\textwidth]{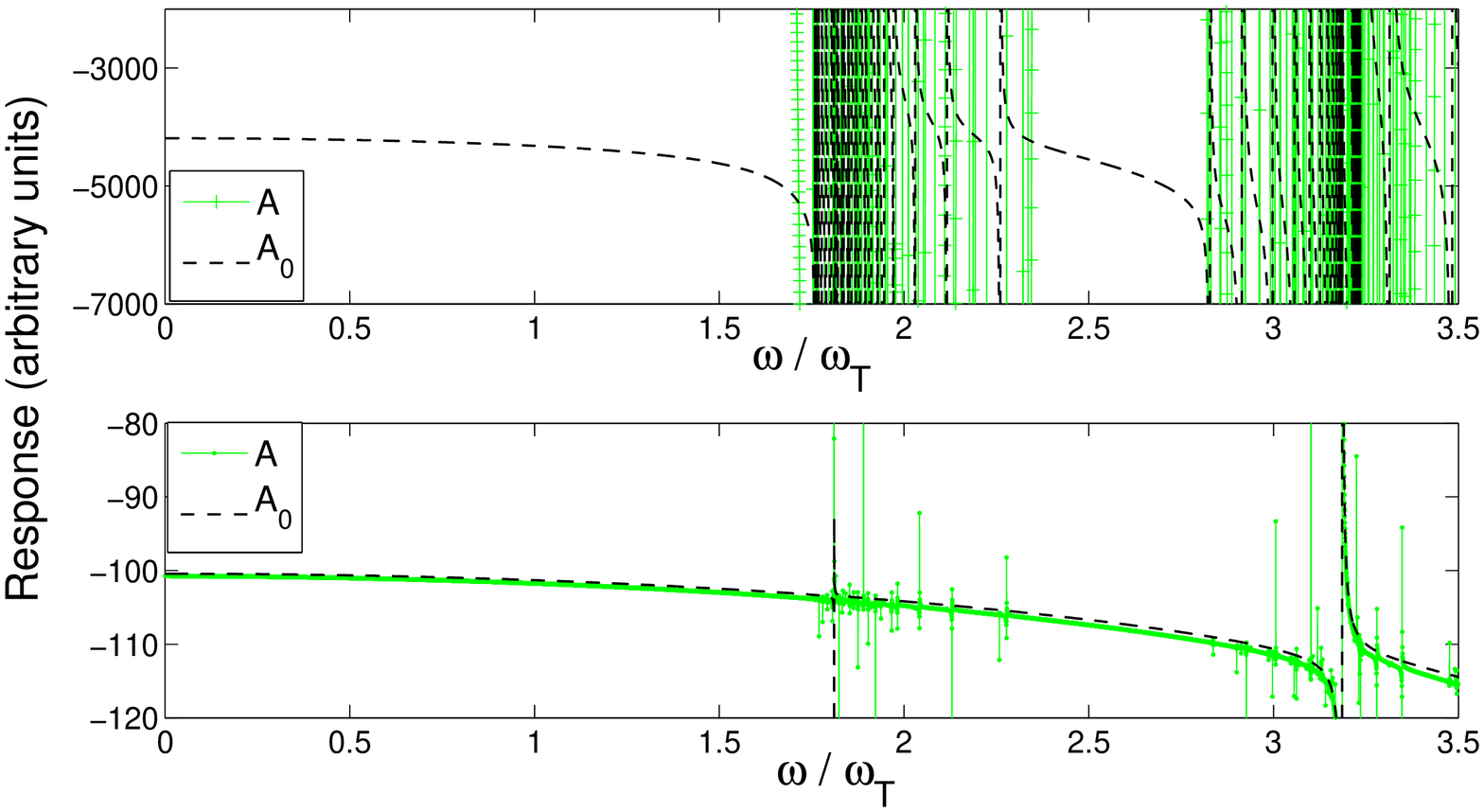}
  \caption{(Color online) The density response determined using the strength function in
Eq.~\eqref{eq:strengthfunc} (upper figure) produces the same qualitative 
features as the definition used in this work in which the response 
is considered only at the centre of the trap (lower figure). 
Here $k_\mr{F}a = -0.84$ and $\Delta(0)=3.9\,\hbar\omega_\mr{T}$.}
  \label{fig:strengthfunc}
\end{figure}

In a trapped and nonuniform system the density response depends on
the positions at which the perturbation is applied on and where the 
response is measured at. In this work we have considered the response at the
centre of the trap, i.e. we defined 
\begin{equation}
  \mathcal{A}(\omega) = \left. \mathcal{A}_{\uparrow\uparrow}(r_1=0,r_5=0,\omega)\right|_{\phi_{\uparrow},L=0}.
\label{eq:localfunc}
\end{equation}
The reason for this choice is the simplicity and the numerical efficiency,
allowing the calculation of plots such as Fig.~\ref{fig:closed_shell}.
In contrast, the standard measure of the density response is the strength
function, which is obtained by integrating the density response over the
whole trap
\begin{equation}
  \mathcal{A}^{\mr S}(\omega) = \int d{\bf r_1} d{\bf r_5} \left. \mathcal{A}_{\uparrow\uparrow}(r_1,r_5,\omega)\right|_{\phi_{\uparrow},L=0}.
\label{eq:strengthfunc}
\end{equation}
The two definitions, Eqs.~\eqref{eq:localfunc} 
and~\eqref{eq:strengthfunc} produce slightly different results, but the main
features of the full density response are the same: the frequencies 
and the widths of the collective mode bands are the same for both as shown
in Fig.~\ref{fig:strengthfunc}. The reason for the similarities between two
methods is that the generalized random phase approximation couples the 
excitations at the centre of the trap to excitations all over the system,
and hence one can excite for example surface modes even by considering 
only the centre of the trap. Such indirect effects come at the cost of
reducing the amplitude of the corresponding collective mode peaks.
However, when considering only the frequencies and the widths of 
the collective excitation bands, the actual amplitudes of various peaks 
are irrelevant. 

In contrast, looking only at single particle excitations $\mathcal{A}_0$,
the local response (at the centre of the trap) has only three low energy
branches, whereas the 'single-particle' strength function 
$\mathcal{A}_0^{\mr S} = \int d{\bf r_1} d{\bf r_5} \mathcal{A}_0(r_1,r_5,\omega)$
has a large number of peaks, corresponding to single-particle excitations
at different parts of the trap. Without the coupling provided by the
generalized random phase approximation, the single-particle excitations 
localized at the centre of the trap will remain localized. The single
particle excitation spectra are thus clearly different in the two approaches.
However, the experimentally relevant quantity is the full density response. 

\begin{figure}
  \includegraphics[width=0.45\textwidth]{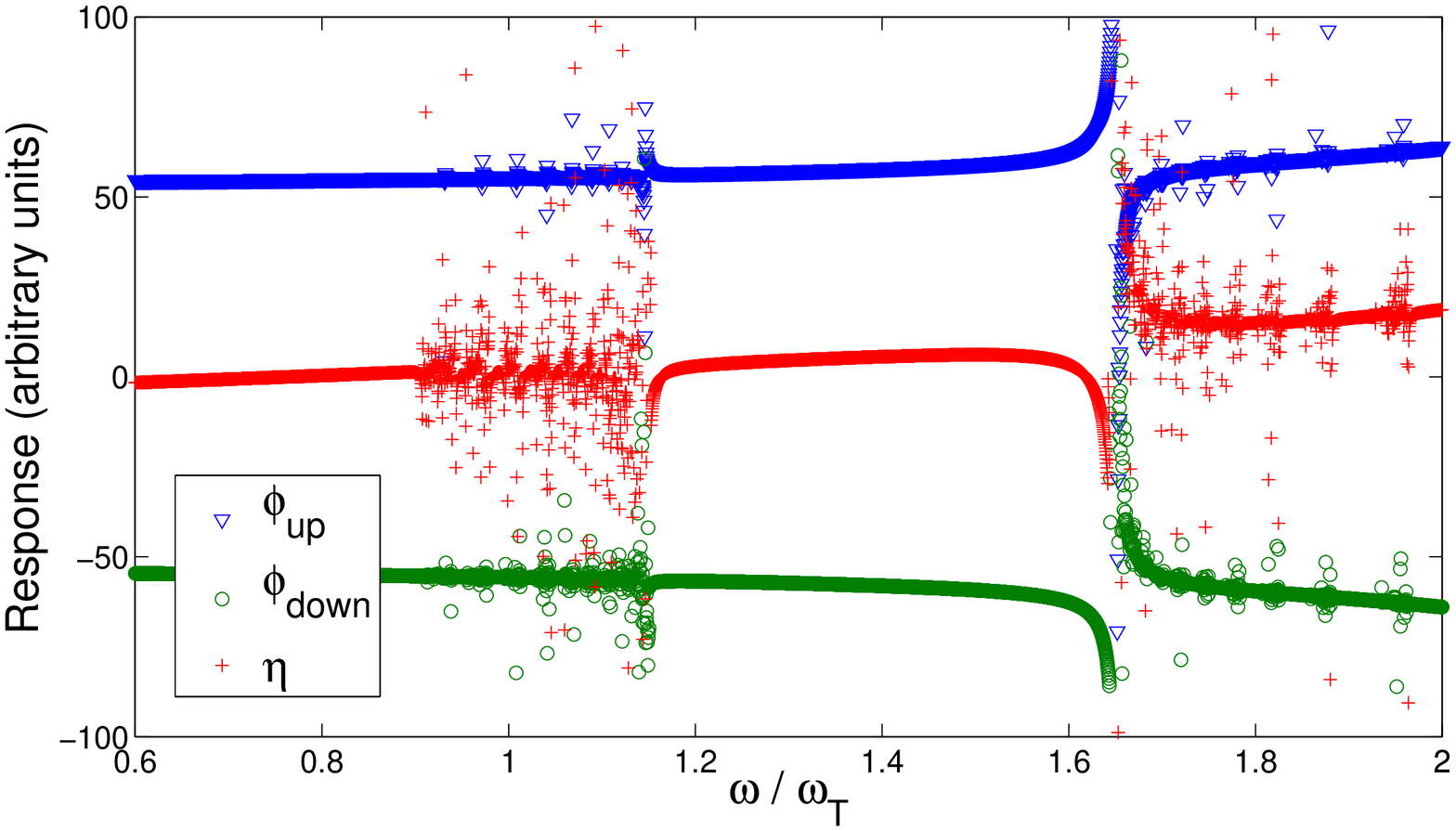}
  \caption{(Color online) The density responses $\mathcal{A}(\omega)$ for different probing fields $\phi_{\uparrow}$, $\phi_{\downarrow}$ and $\eta$ for $k_\mr{F}a=-0.61$ produce
the same collective excitation bands.}
  \label{fig:12}
\end{figure}

The density response results discussed in this manuscript have considered
only the response to probing field $\phi_\uparrow$ directly affecting
only the density of spin-$\uparrow$ atoms. Fig.~\ref{fig:12} shows the
response for all three probing fields $\phi_{\uparrow}$, $\phi_{\downarrow}$ and
$\eta$, as described in Eq.~\eqref{eq:Perturbation}.  While the magnitudes
of the individual peaks are different for different probing fields, the 
main features of the responses are the same, namely the
frequencies and the widths of the collective excitation bands. The same
is true also when considering the response of the pairing field.

\section{Discussion}
\label{sec:discussion}

In this work, we have used the random phase approximation together with the
self-consistent Bogoliubov-deGennes method for studying the density response
of a trapped Fermi gas. Concentrating on the monopole mode in a spherically
symmetric trap we have analysed in detail the interesting crossover regime
in which the collisionless gas becomes strongly interacting. We observe the merging
of a pair vibration mode, originating from a low frequency excitation in the
extreme weakly interacting gas, and a collisionless hydrodynamic mode 
at frequency $2\,\omega_\mr{T}$,
the combined collective mode eventually becoming the strongly 
interacting hydrodynamic Goldstone mode
in the unitary regime. The merging of the two collective mode branches is 
signalled by a depression of the mode frequencies and an increase in the damping
rate, in good agreement with the experiments done in elongated traps.  In the future 
it will be interesting to generalise the method to non-spherically
symmetric systems~\cite{Tezuka2010a,Baksmaty2011a,Kim2011a,Baksmaty2011b}, 
allowing the study of various radial and axial modes and thus a more 
detailed comparison with the reported and possibly new experiments. 
Other interesting
future extensions of the method are studies of the FFLO state~\cite{Edge2009a,Heikkinen2011a}, sensor applications~\cite{Koponen2009a} and dimensionality 
effects~\cite{Koponen2006a}.

\section*{Acknowledgements}
We would like to acknowledge useful discussions with M.O.J. Heikkinen.
This work was supported by National Graduate School in Materials Physics
and the Academy of Finland (Project Nos. 210953, 213362, 217043, 217045, 
135000, and 141039), conducted as a part of a EURYI scheme grant, 
see www.esf.org/euryi, and supported in part by the National 
Science Foundation under Grant No. PHY05-51164.

\bibliographystyle{apsrev4-1}
\bibliography{monopole_bdg_rpa}

\appendix

\section{Density response within the Random Phase Approximation}
\label{sec:Density-response-within}

In this appendix we discuss the derivation of the density response 
Eq.~\eqref{eq:Main}, starting from the Hamiltonian in 
Eq.~\eqref{eq:Hamiltonian}. The Green's function 
$\widehat{G}(\mathbf{1},\mathbf{2})$ 
satisfies the following equation
\begin{widetext}
\begin{equation}
\begin{split}
  &\left(-\frac{\partial}{\partial\tau_{1}}+\left(\frac{\nabla_1^2}{2m}+\mu-\frac{m\omega_{\mr T}^2x_1^2}{2} \right)\tau_3\right)\widehat{G}({\bf 1,2})=
  \hat{I}\delta({\bf 1-2})+\hat W({\bf 1}) \widehat{G}({\bf 1,2})+ g_0 \hat M_\mr{int}({\bf 1,2}),
    \label{eq:app-3}
\end{split}
\end{equation}
\end{widetext}
where $\hat I$ is the identity operator,
\begin{equation}
  \hat W(\bf{1})=\left(\begin{array}{cc}
  -e\phi_{\uparrow}({\bf 1}) & \eta^{*}({\bf 1})\\
  \eta(\bf{1}) & e\phi_\downarrow({\bf 1})\end{array}\right),
\end{equation}
and 
\begin{equation}
\begin{split}
  & \hat{M}_\mr{int}({\bf 1,2})  =  \\
  & = \left(\begin{array}{cc}
    \Bigl\langle T\psi_\uparrow^\dagger(\mathbf{2})n(\mathbf{1})\psi_\uparrow(\mathbf{1})\Bigr\rangle & \Bigl\langle T\psi_{\downarrow}(\mathbf{2})n(\mathbf{1})\psi_{\uparrow}(\mathbf{1})\Bigr\rangle\\
    -\Bigl\langle T\psi_{\uparrow}^{\dagger}(\mathbf{2})\psi_{\downarrow}^{\dagger}(\mathbf{1})n(\mathbf{1})\Bigr\rangle & \Bigl\langle T\psi_{\downarrow}^{\dagger}(\mathbf{1})n(\mathbf{1})\psi_{\downarrow}(\mathbf{2})\Bigr\rangle\end{array}\right),
\end{split}
\end{equation}
where $\Bigl\langle T\ldots\Bigr\rangle$ is a time-ordered correlator.
The bare Green's function $\widehat{G}_{0}(\mathbf{1},\mathbf{2})$ 
is defined similarly but in the absence of the perturbation and 
the interactions $g_0=0$
\begin{equation}
  \left(-\frac{\partial}{\partial\tau_{1}}+\left(\frac{\nabla_{1}^{2}}{2m}+\mu-\frac{m\omega_{\mr T}^{2}x_{1}^{2}}{2}\right)\tau_{3}\right)\widehat{G}_{0}(\mathbf{1},\mathbf{2})=\hat{I}\delta(\mathbf{1}-\mathbf{2}).
  \label{eq:app-4}
\end{equation}
Now, we can express Eq.~\eqref{eq:app-3} using $\widehat{G}_{0}(\mathbf{1},\mathbf{2})$ as follows
\begin{equation}
\begin{split}
  \widehat{G}(\mathbf{1},\mathbf{2})&=\widehat{G}_{0}(\mathbf{1},\mathbf{2})\\
  &+\int d\mathbf{3}\int d\mathbf{4}\widehat{G}_{0}(\mathbf{1},\mathbf{3})\hat W(\mathbf{3})\delta(\mathbf{3}-\mathbf{4})\widehat{G}(\mathbf{4},\mathbf{2})\\
  &+g_{0}\int d\mathbf{3}\int d\mathbf{4}\widehat{G}_{0}(\mathbf{1},\mathbf{3})\hat \Sigma(\mathbf{3},\mathbf{4})\widehat{G}(\mathbf{4},\mathbf{2}),
  \label{eq:app-5}
\end{split}
\end{equation}
where the self-energy $\hat \Sigma(\mathbf{3},\mathbf{4})= \int d\mathbf{5}  \hat{M}_\mr{int}({\bf 3,5}) \widehat{G}^{-1}({\bf 5,4})$. Eq.~\eqref{eq:app-5}
can be formally written as
\begin{equation}
  \widehat{G}^{-1}(\mathbf{1},\mathbf{2})=\widehat{G}_{0}^{-1}(\mathbf{1},\mathbf{2})-\hat W(\mathbf{1})\delta(\mathbf{1}-\mathbf{2})-\hat \Sigma(\mathbf{1},\mathbf{2}).
  \label{eq:app-6}
\end{equation}

The next step is to apply the variational derivative $\frac{\delta}{\delta h(\mathbf{5})}$ to both sides of Eq.~\eqref{eq:app-6} for separate cases $h=\phi_{\uparrow};\quad\phi_{\downarrow};\quad\eta$.
Applying $\frac{\delta\widehat{G}(\mathbf{1},\mathbf{2})}{\delta h(\mathbf{5})}=-\int d\mathbf{3}d\mathbf{4}\widehat{G}(\mathbf{1},\mathbf{3})\frac{\delta\widehat{G}^{-1}(\mathbf{3},\mathbf{4})}{\delta h(\mathbf{5})}\widehat{G}(\mathbf{4},\mathbf{2})$ yields the equation
\begin{equation}
\begin{split}
 & \frac{\delta\widehat{G}(\mathbf{1},\mathbf{2})}{\delta h(\mathbf{5})}=\hat{A}_{0}(\mathbf{1},\mathbf{2},\mathbf{5})\\
  &+\int d\mathbf{3}\widehat{G}(\mathbf{1},\mathbf{3})\left(-\tau_{3}(-\frac{\delta n(\mathbf{3})}{\delta h(\mathbf{5})}+\frac{\delta\widehat{G}(\mathbf{3},\mathbf{3})}{\delta h(\mathbf{5})}\tau_{3})\right)\widehat{G}(\mathbf{3},\mathbf{2}).
  \label{eq:app-7}
\end{split}
\end{equation}
Here $\hat{A}_{0}(\mathbf{1},\mathbf{2},\mathbf{5})$ depends from
the field $h$, and is $\widehat{G}(\mathbf{1},\mathbf{5})\left(\begin{array}{cc}
  1 & 0\\
  0 & 0\end{array}\right)\widehat{G}(\mathbf{5},\mathbf{2})$, $\widehat{G}(\mathbf{1},\mathbf{5})\left(\begin{array}{cc}
    0 & 0\\
    0 & 1\end{array}\right)\widehat{G}(\mathbf{5},\mathbf{2}),$ $\widehat{G}(\mathbf{1},\mathbf{5})\left(\begin{array}{cc}
      0 & 1\\
      1 & 0\end{array}\right)\widehat{G}(\mathbf{5},\mathbf{2})$ for $h$ equal to $\phi_{\uparrow}$, $\phi_{\downarrow}$ and $\eta$, respectively. 

After transforming $\tilde{G}(\mathbf{1},\mathbf{3})=\tau_{3}\widehat{G}(\mathbf{1},\mathbf{3})$, Eq.~\eqref{eq:app-7} becomes:
\begin{equation}
\begin{split}
  \tilde{A}_{ij}(\mathbf{1},\mathbf{2},\mathbf{5})&=\tilde{A}_{0ij}(\mathbf{1},\mathbf{2},\mathbf{5})\\
  &+g_{0}\int d\mathbf{3}L_{ikkj}(\mathbf{1},\mathbf{2},\mathbf{3})\tilde{A}_{ll}(\mathbf{3},\mathbf{3},\mathbf{5})\\
  &-g_{0}\int d\mathbf{3}L_{iklj}(\mathbf{1},\mathbf{2},\mathbf{3})\tilde{A}_{kl}(\mathbf{3},\mathbf{3},\mathbf{5}),
  \label{eq:app-8}
\end{split}
\end{equation}
where $\tilde{A}_{ij}(\mathbf{1},\mathbf{2},\mathbf{5})=\frac{\delta\tilde{G}_{ij}(\mathbf{1},\mathbf{2})}{\delta h(\mathbf{5})}$,
$L_{iklj}(\mathbf{1},\mathbf{2},\mathbf{3})=\tilde{G}_{ik}(\mathbf{1},\mathbf{3})\tilde{G_{lj}}(\mathbf{3},\mathbf{2})$
and $\tilde{A}_{0}(\mathbf{1},\mathbf{2},\mathbf{5})=\tilde{G}(\mathbf{1},\mathbf{5})\left(\begin{array}{cc}
1 & 0\\
0 & 0\end{array}\right)\tilde{G}(\mathbf{5},\mathbf{2})$, $\tilde{G}(\mathbf{1},\mathbf{5})\left(\begin{array}{cc}
0 & 0\\
0 & 1\end{array}\right)\tilde{G}(\mathbf{5},\mathbf{2})$, $\tilde{G}(\mathbf{1},\mathbf{5})\left(\begin{array}{cc}
0 & 1\\
1 & 0\end{array}\right)\tilde{G}(\mathbf{5},\mathbf{2})$ for $h=\phi_{\uparrow}$, $\phi_{\downarrow}$ and $\eta$, respectively.

\section{Coefficients $\mathcal{L}_{iklj,L}(r_{1},r_{3},\omega)$ via BdG Green's functions}
\label{sec:Coefficients--via}

The Fourier transform of the matrix element $L_{iklj}$ is $L_{iklj}(\mathbf{r}_{1},\mathbf{r}_{3},\omega)=\frac{1}{\beta\hbar}\sum_{\Omega_{n}}G_{ik}(\mathbf{r}_{1},\mathbf{r}_{3},\Omega_{n})G_{lj}(\mathbf{r}_{3},\mathbf{r}_{1},\omega+\Omega_{n})$,
where $\beta=\frac{1}{k_{B}T}$ is the temperature, and $\Omega_{n}=\frac{(2n+1)\pi}{\beta\hbar}$
are Matsubara frequencies. 
As discussed in Section~\ref{sub:Symmetry-of-angular}, the spherical symmetry
allows a great simplification of the problem, and one needs only to 
solve the coefficients $\mathcal{L}_{iklj,L}(r_{1},r_{3},\omega)=\int d\gamma_{13}\sin\gamma_{13}L_{iklj}(\mathbf{r}_{1},\mathbf{r}_{3},\omega)P_{L}(\cos\gamma_{13})$, which can be written as
\begin{equation}
\begin{split}
  &\mathcal{L}_{iklj,L}(r_{1},r_{3},\omega)=\left(2L+1\right)\frac{1}{\beta\hbar}\sum_{L_{1}L_{2}} \left(\begin{array}{ccc}
    L & L_{1} & L_{2}\\
    0 & 0 & 0\end{array}\right)^{2} \times \\
  &\sum_{\Omega_{n}}G_{ik,L_{1}}(r_{1},r_{3},\Omega_{n})G_{lj,L_{2}}(r_{3},r_{1},\omega+\Omega_{n}),
  \label{eq:app-9}
\end{split}
\end{equation}
where
$\left(\begin{array}{ccc}
    L & L_{1} & L_{2}\\
    0 & 0 & 0\end{array}\right)$
are Wigner 3j coefficients.

The Bogoliubov-deGennes Green's functions are provided by Eq.~\eqref{eq:a-3},
yielding the final result
\begin{widetext}
\begin{equation}
\begin{split}
&\mathcal{L}_{iklj,L}(r_{1},r_{3},\omega)=\left(2L+1\right)\sum_{L_{1}L_{2}}\left(\begin{array}{ccc}
L & L_{1} & L_{2}\\
0 & 0 & 0\end{array}\right)^{2}\frac{2L_{1}+1}{4\pi}\frac{2L_{2}+1}{4\pi}\\
&\times \sum_{J_{1}J_{2}}\left(\lambda_{J_{1}L_{1},ik}^{-}\lambda_{J_{2}L_{2},lj}^{-}\frac{n_{F}(E_{J_{1}L_{1}})-n_{F}(E_{J_{2}L_{2}})}{\omega+E_{J_{1}L_{1}}-E_{J_{2}L_{2}}}+\lambda_{J_{1}L_{1},ik}^{+}\lambda_{J_{2}L_{2},lj}^{+}\frac{n_{F}(-E_{J_{1}L_{1}})-n_{F}(-E_{J_{2}L_{2}})}{\omega-E_{J_{1}L_{1}}+E_{J_{2}L_{2}}}\right. \\
&+\left.\lambda_{J_{1}L_{1},ik}^{-}\lambda_{J_{2}L_{2},lj}^{+}\frac{n_{F}(E_{J_{1}L_{1}})-n_{F}(-E_{J_{2}L_{2}})}{\omega+E_{J_{1}L_{1}}+E_{J_{2}L_{2}}}+\lambda_{J_{1}L_{1},ik}^{+}\lambda_{J_{2}L_{2},lj}^{-}\frac{n_{F}(-E_{J_{1}L_{1}})-n_{F}(E_{J_{2}L_{2}})}{\omega-E_{J_{1}L_{1}}-E_{J_{2}L_{2}}}\right),
\label{eq:a-4-1}
\end{split}
\end{equation}
\end{widetext}
where $\lambda_{J_{1}L_{1},ik}^{\pm}=\Lambda_{J_{1}L_{1},i}^{\pm}(r_{1})\Lambda_{J_{1}L_{1},k}^{\pm\dagger}(r_{3})$, $\lambda_{J_{2}L_{2},lj}^{\pm}=\Lambda_{J_{2}L_{2},l}^{\pm}(r_{3})\Lambda_{J_{2}L_{2},j}^{\pm\dagger}(r_{1})$, and the quasiparticle wavefunctions
$\Lambda^{\pm}$ are defined in the main text below Eq.~\eqref{eq:a-3}.

\section{Procedure of defining the peaks}
\label{sub:Procedure-of-defining}

Numerical calculations with very high precision show that the peaks 
in the density response have the shape 
$\sim\frac{1}{(\omega-\omega_{0})^n}$ where $\omega_{0}$ is the 
(yet unknown) frequency of the mode. However, for most of the figures
in this work, we have limited the frequency grid resolution to
$\delta \omega = 3\cdot 10^{-4}\,\omega_{\mr T}$. This resolution is high enough
to show the existence of a peak but not to show the detailed shape.
For this resolution, some modes result in extremely high peaks (if the
frequency grid happened to coincide with the mode frequency) regardless
of the actual prefactor or the amplitude. We decided to define 'a peak'
as a frequency for which the derivative of the density response is higher than
some chosen cut-off. The reason for using the derivative instead of the
height of the peak is that the base level of the response (the value 
of the response in areas away from any peaks) is slowly changing with 
increasing frequency $\omega$. Such slow changes in the base level
do not affect the derivative and the peaks are more clearly seen. The caveat
is the higher sensitivity to numerical noise.
For creating Figs.~\ref{fig:closed_shell}
and~\ref{fig:open_shell} we mark a peak when the derivative 
$\frac{\partial A(\omega)}{\partial \omega} > 3 \cdot 10^3$. 
The choice of this value is a 
compromise between limiting the required resolution and in avoiding the 
effect of the numerical noise. The main results, such as the frequencies 
of the collective modes or the widths of the excitation bands, are not 
sensitive to this choice.

\end{document}